\documentclass[prd, aps, superscriptaddress, preprintnumbers, twocolumn, floatfix, nofootinbib]{revtex4}

\usepackage{amsmath}    
\usepackage{graphicx}   
\usepackage{verbatim}   
\usepackage{color}      
\usepackage{subfigure}  
\usepackage{hyperref}   
\usepackage{float}     
\usepackage{mdwlist}

\def\be{\begin{equation}}
\def\ee{\end{equation}}

\begin{document}

\title{Dark Matter Distribution Induced by a Cosmic String Wake in the Nonlinear Regime}

\author{Disrael Camargo Neves da Cunha\footnote{camargod@hep.physics.mcgill.ca}}
\affiliation{Department of Physics, McGill University,Montreal, QC, H3A 2T8, Canada }
\author{Joachim Harnois-Deraps\footnote{Joachim's email}}
\affiliation{Scottish Universities Physics Alliance, Institute for Astronomy, University of Edinburgh, Blackford Hill, Scotland, UK}
\author{Robert Brandenberger\footnote{rhb@physics.mcgill.ca}}
\affiliation{Department of Physics, McGill University,Montreal, QC, H3A 2T8, Canada }
\author{Adam Amara\footnote{adam.amara@phys.ethz.ch}}
\affiliation{ETH Zurich, Department of Physics, Wolfgang-Pauli-Strasse 27, 8093 Zurich, Switzerland} 
\author{Alexandre Refregier\footnote{alexandre.refregier@phys.ethz.ch}}
\affiliation{ETH Zurich, Department of Physics, Wolfgang-Pauli-Strasse 27, 8093 Zurich, Switzerland} 
\date{\today}

\begin{abstract}

We study the distribution of dark matter in the nonlinear regime in a model in
which the primordial fluctuations include, in addition to the dominant primordial Gaussian
fluctuations generated by the standard $\Lambda CDM$ cosmological model,
the effects of a cosmic string wake set up at the time of equal matter and radiation,
making use of cosmological $N$-body simulations. At early
times the string wake leads to a planar overdensity of dark matter. We study
how this non-Gaussian pattern of a cosmic string wake evolves in the presence of the
Gaussian perturbations, making use of wavelet and ridgelet-like statistics specifically designed to extract
string wake signals. At late times the Gaussian fluctuations disrupt the string wake.
We find that for a string tension of $G \mu = 10^{-7}$, a value just below the current
observational limit, the effects of a string wake can be identified in the dark matter distribution, using the
current level of the statistical analysis, down to a redshift of $z = 10$.

\end{abstract}

\maketitle

\tableofcontents

\section{Introduction}

Cosmic strings are linear topological defects which arise in a large class of quantum field
theory models describing physics beyond the Standard Model. If Nature is described
by such a model, then a network of strings inevitably forms in the early universe and
persists to the present time \cite{Kibble}. Strings are thin lines of trapped energy density
and their gravitational effects lead to specific signatures which can be searched for
in cosmological observations.

The network of cosmic strings \footnote{We are focusing on non-superconducting
strings. In some quantum field theory models the strings can be superconducting \cite{Witten}
which will lead to additional effects of non-gravitational origin.} which form in a gauge field theory 
approaches a {\it scaling solution} in which the statistical properties of the distribution
of strings is independent of time when all lengths are scaled by the Hubble radius (see
e.g. \cite{VS, HK, RHBCSrev} for reviews of cosmic strings and their role in cosmology). The
string distribution has two components: firstly a network of infinite strings with mean
curvature radius comparable to the Hubble radius $t$ (where $t$ is the physical
time), and secondly a distribution of string loops with radii $R < t$. The loops result
from the long string intersections which also are responsible for maintaining the
long string scaling distribution. Analytical arguments lead to the conclusion that
the number $N$ of long string segments that pass through any Hubble volume is
of the order $1$. The exact number must be determined in numerical cosmic string
evolution simulations (see e.g. \cite{CSsimuls} for some recent results). Current
evidence is that $N \sim 10$.
 
Cosmic strings are characterized by their mass per unit length $\mu$ which is usually
expressed in terms of the dimensionless quantity $G \mu$ (where $G$ is Newton's
gravitational constant \footnote{Note that we are using natural units in which the
speed of light is $c = 1$.}). The value of $\mu$ is determined by the energy scale
$\eta$ at which the strings are formed (it is the energy scale of the phase 
transition leading to the strings)
\be
\mu \, \sim \, \eta^2 \, .
\ee
The strength of the signatures of cosmic strings in the sky are proportional to $G \mu$. 
The current upper bound on the string tension is
\be \label{CSbound}
G \mu \, < \, 1.5 \times 10^{-7} \,
\ee
and is derived from the features of the angular power of cosmic microwave background
(CMB) anisotropies \cite{Dvorkin, PlanckCS} (see also \cite{CMBlimits} for
some older works). Searching for cosmic strings in the
sky hence is a way to probe particle physics beyond the Standard Model ``from
top down'', as opposed to accelerator searches which are more sensitive to low
values of $\eta$ and hence probe particle physics ``from bottom up''. The current
bound on $G \mu$ already rules out the class of ``Grand Unified'' particle physics models
containing cosmic string solutions with a scale of symmetry breaking which is on
the upper end of the preferred range. Improving the upper bound on the cosmic
string tension will lead to tighter constraints on particle physics models (see e.g.
\cite{RHBtopRev} for an elaboration on these points).

The long string segments lead to non-Gaussian signals in the sky characterized
by specific geometrical signatures in position space maps. One set of string
signatures comes from lensing produced by a string. Space perpendicular to
a long straight string segment is a cone with deficit angle $\alpha$ given by \cite{deficit}
\be \label{defangle}
\alpha \, = \, 8 \pi G \mu \, .
\ee
The deficit angle extends to a distance $t$ from the string \cite{Joao}.
Cosmic strings are relativistic objects and hence the curvature of the string segments
(the curvature radius is of the order $t$) will induce relativistic motion of the string
in the plane perpendicular to the tangent vector of the string. This will lead to line 
discontinuities in CMB anisotropy maps \cite{KS} of magnitude
\be
\frac{\delta T}{T} \, = \, 4 \pi v_s \gamma_s G \mu \, ,
\ee
where $v_s$ is the transverse speed of the string segment, and $\gamma_s$ is
the corresponding relativistic gamma factor. 

A moving long string segment will also induce a velocity
perturbation behind the string towards to plane determined by the tangent vector of
the string and the velocity vector. This leads to a region behind the string with
twice the background density called a cosmic string {\it wake} \cite{wake}. A wake 
produced by a string passing through matter at time $t$ will have
comoving planar dimensions given by the Hubble radius at time $t$, and 
a comoving thickness which initially is given by the deficit angle (\ref{defangle}) times
the Hubble radius and 
which grows linearly in time as given by the result of an analytical analysis 
\cite{wakegrowth} making use of the Zel'dovich approximation \cite{Zeld}. Hence,
the comoving dimensions of a wake produced at time $t$ are
\be
c_1 t z(t) \times t z(t) v_s \gamma_s \times 4 \pi G \mu v_s \gamma_s z(t) t \, ,
\ee
where $z(t)$ is the redshift at time $t$, and $c_1$ is a constant of order unity
which gives the string curvature radius relative to $t$.

According to the cosmic string network scaling solution, strings lead to
a set of line discontinuities in CMB temperature maps. The overall
distribution of these discontinuities is scale-invariant. However, since
cosmic strings are primordial isocurvature fluctuations, they do not give
rise to acoustic oscillations in the angular CMB temperature power spectrum,
oscillations which are typical of adiabatic perturbations \cite{noacoustic}. Hence, detailed
measurements of the CMB angular power spectrum leads to the
constraints on the string tension is given by (\ref{CSbound}). It is
likely that the bound can be strengthened by analyzing CMB temperature
maps in position space using statistical methods designed to identify
linear discontinuities. Initial studies using the Canny edge detection algorithms
\cite{Danos}, wavelets \cite{Hergt, PlanckCS, Hiranya},  curvelets \cite{Hergt}
and machine learning tools \cite{Oscar2}
have shown that good angular resolution is key to obtaining improved constaints
\footnote{See also \cite{Smoot, Wright} for earlier searches for position space
signals of cosmic strings in CMB temperature maps.}. Cosmic
strings also lead to direct B-mode polarization in the CMB sky \cite{Holder1}
(see also \cite{RHBCSrevNew} for a recent review of signatures of cosmic
strings in new observational windows) \footnote{Searching for cosmic strings in
position space has an additional advantage over analyzing only correlation
functions such as the power spectrum: searching for signals of individual strings
in position space maps leads to less sensitivity to the parameter $N$ (number
of long strings per Hubble volume) which is not yet well determined).}. 

Cosmic strings also lead to distinct patterns in 21cm redshift surveys: a
cosmic string wake present at a redshift before reionization will lead to
a three-dimensional wedge of extra absorption in the 21cm maps because
at these redshifts the wake is a region of twice the background density of neutral
hydrogen and CMB photons passing through a wake suffer twice the absorption
compared to photons which do not pass through the wake \cite{Holder2}. Strings
also lead to a Wouthuysen-Field brightness trough in the integrated 21cm signal
\cite{Oscar}.

In contrast, there has been little recent work on how well the cosmic string
tension can be constrained by the large-scale structure of the Universe at lower
redshifts, well into the nonlinear region of gravitational clustering \footnote{Most previous
work on cosmic strings signals in the large-scale structure has been in the context of 
string models \cite{CSearly} without $\Lambda CDM$ fluctuations.}. In this
paper we take first steps at studying these signals. We will study how well a
single cosmic string wake can be identified in $N$-body dark matter simulations of gravitational
clustering. 

Specifically, we include the effects of a cosmic string wake in a cosmological $N$-body
simulation which evolves the dark matter distribution. We introduce a statistic which is
designed to pick out the signal of a cosmic string wake in the ``noise'' of the primordial
Gaussian fluctuations in a $\Lambda CDM$ cosmology. Since the string wake grows
only in direction perpendicular to the plane of the wake, whereas the Gaussian fluctuations
grow in all three dimensions, the Gaussian fluctuations will eventually disrupt the wake,
as studied analytically in \cite{us}. However, even once the wake has been
locally disrupted, its global signal will persist for some time. We study how the redshift when 
this global signal ceases to be identifiable varies as the string tension changes.
In the following, we shall call the Gaussian fluctuations in a $\Lambda CDM$ cosmology
simply as ``Gaussian noise''.

\section{Wake Disruption}

The challenge of identifying cosmic string wake signals in the nonlinear regime of
structure formation was addressed in \cite{us}. If a cosmic string wake is added to
the initial conditions of a cosmological model which is characterized by a scale-invariant
spectrum of primordial Gaussian cosmological perturbations, then the wake is clearly identifiable
at high redshifts since the Gaussian perturbations are all in the linear regime whereas
the wake is already a nonlinear density contrast. However, once the Gaussian perturbations
become nonlinear they will start to disrupt the wake.

A first criterion for the stability of a wake is that the local displacement $S_{k}$
of matter on the comoving scale of the wake thickness $k^{-1}(t)$ due to the Gaussian fluctuations
be smaller than the physical width of the wake $h(t)$, i.e.
\be
S_{k}(t) \, < \, h(t) \, .
\ee
If this condition is satisfied, then the wake should persist as a locally coherent entity.
This condition was called the {\it local stability condition}. A closely related condition
is the {\it local delta condition} which demands that the mean fluctuation $\Delta$ due to
the Gaussian fluctuations on the scale $k$ of the wake thickness be smaller than unity, i.e.
\be \label{local}
\Delta(k, t) \, < \, 1 \, .
\ee
For a string tension of $G\mu = 10^{-7}$ it was found that the local delta condition
is satisfied down to a redshift of $z \simeq 5$. The limiting redshift increases as the
string tension decreases. For $G\mu = 10^{-11}$ the limiting redshift is $z \sim 11$,
and this limiting redshift increases only slowly as the string tension is reduced
further (for a string tension of $G\mu = 10^{-14}$ the limiting redshift is $z \sim 20$).

The above result shows that in principle very high redshift surveys of the distribution
of matter in the universe such as what can be achieved by high redshift 21cm maps
yield a very promising avenue to detect cosmic strings \cite{Holder2}. The challenge,
however, is to be able to identify the very thin features (in redshift direction) which
string wakes will produce.

As was also studied in \cite{us}, wakes might be identifiable through the global
mass distribution which they induce even if they are locally disrupted. We can
ask the question whether the Gaussian fluctuations are able to induce a nonlinear
overdensity in a box of the expected dimensions of a string wake. The contribution
of the Gaussian fluctuations to the variance in a such a box $B(G\mu)$
is
\be
\sigma_w^2 \, = \, \frac{g^2(z)}{(2 \pi)^3} \int d^3k P(|k|) W_w^2(k) \, ,
\ee
where $P(|k|)$ is the power spectrum of the Gaussian noise, $g(z)$ is the cosmological
growth factor, and $W_w(k)$ is a non-isotropic window function which filters out
contributions from modes which have wavelength smaller than the width of the
wake in one direction, and smaller than the extent of the wake in the two other
dimensions. If $\sigma_w < 1$, then a string wake can be identified by its
global signal (it will produce a nonlinear density contrast in this box). Thus,
we can define the {\it global delta condition} for the identifiability of a string
wake:
\be \label{global}
\sigma_w \, < \, 1 \, .
\ee

It was found \cite{us} that for a roughly scale-invariant power spectrum of primordial
fluctuations the result for $\sigma_w$ is to first order independent of the thickness
of the wake, and that the condition (\ref{global}) remains satisfied down to redshift $0$.
Hence, in the absence of noise and with unlimited resolution a string wake should
be identifiable even at the present time for any value of $G\mu$. In practice, however,
the limited resolution of a survey (and the limiting resolution of numerical simulations)
will limit the redshift range where the string wake can be detected.

The goal of the present study is to determine to what value of $G\mu$ cosmic
string wakes can be identified in practice. Ultimately we are interested in comparing
the results of numerical simulations of the distribution of matter, obtained if the
usual initial conditions for the primordial fluctuations are supplemented with the
presence of a cosmic string wake, with observational data.  In the current project
we will study the distribution of dark matter only. Any observational data set will
have a limiting resolution in the same way that any numerical simulation has a
resolution limit. These limits will render the effects of string wakes harder
and harder to detect the smaller the value of $G\mu$, in spite of the fact that
the result (\ref{global}) is independent of $G\mu$. In this paper we wish to study 
whether the wake of a string with tension $G\mu = 10^{-7}$, a value just
below the current upper bound, can be identified with simulations having a resolution 
which can currently be achieved. 

In the next section we describe the simulation code and various performance tests
of the code which we have performed. For these test runs we use a value of
$G\mu$ which is larger than the current upper bound in order to better visualize
the results. In Section 4 we then present the output of runs for values of
$G\mu$ down to $G\mu = 10^{-7}$, and study down to which redshift the wake
signal can be identified with various statistics. In Section 5 we summarize the
results and discuss prospects for deriving improved limits.

\section{Simulations}

\subsection{The Code}

This section describes the main features of the $N$-body simulations that we use to model 
numerically the wake evolution and its impact on the density field. 
We detail our wake insertion strategy and validate our results with consistency checks.

The simulations were produced with {\small CUBEP$^3$M}, a public high performance 
cosmological $N$-body code based on a two-level mesh gravity solver
augmented with sub-grid particle-particle interactions \cite{code}.  
The initial conditions generator reads a transfer function constructed with the {\small CAMB} 
online toolkit\footnote{CAMB:https://lambda.gsfc.nasa.gov/toolbox/tb\_camb\_form.cfm} and 
produces $\Lambda$CDM fluctuations at any chosen initial redshift $z_i $ with the following 
cosmological parameter: $\Omega_{\Lambda}=0.7095$, $\Omega_{\rm b}=0.0445$, 
$\Omega_{\rm CDM}=0.246$, $n_{t}=1$, $n_{s}=0.96$, $\sigma_{8}=0.8628$, 
$h=0.70$, $T_{\rm cmb}(t_{0})=2.7255$. The initial 
redshift is chosen such that the initial fluctuations are in the linear regime for all 
scales that we resolve. Except few cases, the intial redshift was $z_i=63$. Particles are then displaced using linear theory \cite{Zeld}, 
then evolved with {\small CUBEP$^3$M} until redshift $z=0$. Several test simulations 
were performed with various computational power on four systems: a laptop with 
4 processors, a 64 cores computer cluster in the McGill Physics Department called irulan,
and a set of 128 cores from the Guillimin Cluster and, finally, a set of 128 cores from the Graham Cluster. The two last clusters are part of the 
Compute Canada Consortium. The first two set of simulations were launched as a single MPI task jobs,
whereas the two last ones were distributed over 8 compute nodes. The cosmological 
volume and the number of particles were varied, as summarized in 
Table \ref{table:nbody_specs}. The most powerful simulation was performed in 
Graham, in a volume of $L_{\rm box} = 64$Mpc/$h$ per side, with 
$nc$ = 2048 cells per dimension (corresponding to 1024 particles per dimension).
The phase space output data was saved for checkpoints chosen in equal spaced 
logarithmic intervals for the scale factor. In addition to that, a few more checkpoints were 
added, corresponding in total to the following redshifts: $63,31,15,10,7,5,4,3,2,1,0.5,0$ (some other simulations contain $255$ and $127$ as well).

\begin{table}[htbp]
   \centering
   \begin{tabular}{@{} ccr @{}} 
 
    \hline
     Machine & $L_{\rm box}$ (in Mpc/$h$)& $nc$ \\
    \hline
     ACER     &  40 & 240  \\
     irulan      & 64 and 32  & 512 and 256 \\
     Guillimin  & 64  &  1024\\
     Graham  & 64  &  1024 and 2048\\
    \hline
 
   \end{tabular}
   \caption{Configuration of the various $N$-body simulations. 
   $L_{\rm box}$ and $nc$ are the side of the cosmological box and number of 
   cells per dimension, respectively.  }
   \label{table:nbody_specs}
\end{table}

\begin{figure}
\includegraphics[height=5cm]{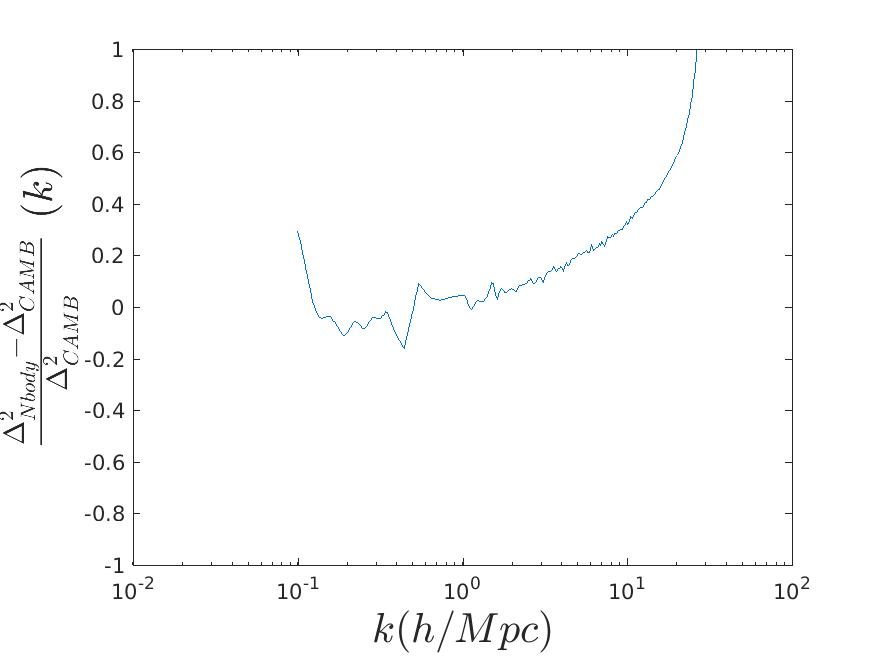}
\caption{Plot of fractional error of the dimensionless power spectrum compared with the HALOFIT predictions of the online CAMB tool} 
\label{fig1}
\end{figure}

The $\Lambda CDM$ part of the $N$-body code has been shown to match the 
predictions to within 5\% over a large range of scales. 
We verify this  in Fig. \ref{fig1}, where we compare the matter power spectrum  
$P(k)$ with the predictions at $z = 15$ for a ACER simulation with 
$L_{\rm box} = 40$Mpc/$h$ and $nc = 240$.
The power spectrum is computed by first assigning the particles onto a density 
grid $\delta(x)$ using the cloud-in-cell interpolation, then squaring the Fourier 
transformed field $\delta(k)$ and averaging over the solid angle $\Omega$: 
$P(k) = \langle |\delta(k)|^2 \rangle_{\Omega}$. The mass assignment scheme has been removed in this calculation, but the shot noise was not removed, which explains the large rise at $k > 10 h/{\rm Mpc}$.
 We observe that the agreement is indeed as expected, with a 10\% match for 
 $k < 3.0h$/Mpc, corresponding to scales 1.04 Mpc/$h$.
 
The reason we cannot achieve $5\%$ precision on $P(k)$ is due to the fact that we do 
not fully capture the linear scales, because we are considering a small box size. We 
would normally need $L_{box} > 200\ {\rm Mpc }/h$ to get $5\%$ precision on $P(k)$. In our 
case this is not an issue, since we are still producing representative $\Lambda CDM$
fluctuations from which the wake must be extracted.

\subsection{Wake Insertion}

One of the goals is to produce particle distributions including the
effects of a wake with a cosmic string tension compatible with the 
current limit of $G\mu = 10^{-7}$, which corresponds to a comoving
width of $\approx 0.003 h^{-1} {\rm Mpc}$ at redshift 20, a redshift in which we 
have confidence that the wake is not yet locally disrupted \cite{us}. 

For a given simulation without a wake, we evolved two independent 
simulations corresponding to wake insertions at redshifts $z = 31$ and 
$z = 15$ to test the sensibility of the results on the time of wake insertion. 
The advantage of an early insertion is that the dark matter distribution is in 
the linear regime. On the other hand a late wake insertion corresponds to 
a thicker wake at the time of insertion, giving a better resolution of its 
thickness.

For each configuration,  a $\Lambda$CDM-only simulation was evolved to 
$z = 0$, without the wake, writing the particle phase space at a number of redshifts
including the wake insertion redshift. We next modified the particle phase space at 
the wake insertion redshift by displacing the particles and also giving them a velocity 
kick towards the central plane. The magnitudes
of the velocity and displacement are calculated according to the Zel'dovich
approximation \cite{Zeld}.

We consider a wake which was laid down at the time $t_{eq}$ of equal matter 
and radiation (such wakes are the most numerous and also the thickest). Their 
comoving planar distance $d$ is given by the comoving horizon at $t_{eq}$, namely
\be
d \, = \, z(t_{eq})^{-1/2} t_0 \, 
\ee
where $t_0$ is the present time. This distance is larger than the size of our
simulation box, which justifies inserting the effects of a wake as a planar
perturbation. The initial velocity perturbation towards the plane of the wake
which the particles receive at $t_{eq}$ is
\be
\delta v \, = \, 4 \pi G \mu v_s \gamma_s \, ,
\ee
where $v_s$ is the transverse velocity of the string and $\gamma_s$ is
the corresponding relativistic gamma factor. Since
cosmic strings are relativistic objects, we will take $v_s \gamma_s = 1/\sqrt{3}$.
In the following equations, however, we leave $v_s \gamma_s$ general.

The initial velocity perturbation leads to a comoving displacement $\psi(t)$
of particles towards the plane of the wake. This problem was discussed in
detail in \cite{wakegrowth}, with the result that the comoving displacement 
at times $t > t_{eq}$ is given by
\be
\psi(t) \, = \, \frac{3}{5} 4 \pi G\mu v_s \gamma_s t_{eq} z(t_{eq}) \frac{z(t_{eq})}{z(t)} \, .
\label{displacement}
\ee
The last factor represents the linear theory growth of the fluctuation, the previous
factor of $z(t_{eq})$ represents the conversion from physical to comoving velocity.
The (comoving) velocity perturbation is
\be
{\dot{\psi}}(t) \, = \, \frac{2}{5} 4 \pi G\mu v_s \gamma_s t_{eq} z(t_{eq}) \frac{z(t_{eq})}{z(t)} \frac{1}{t} \, .
\label{velocity}
\ee

In our simulations the displacement and velocity perturbations towards the plane of
the wake were given by (\ref{displacement}) and (\ref{velocity}), respectively, evaluated
at the time $t = t_i$ of wake insertion.
We then reload this modified  particle data into  {\small CUBEP$^3$M} and let the code 
evolve again to redshift $z=0$.
This method ensures that the differences seen in the late time matter fields are 
caused only by the presence of the wake. The $\Lambda$CDM background is otherwise identical.


To test the wake insertion code, simulations were run with a large cosmic string tension of
$G\mu=4.0\times 10^{-6}$. The following three 
panels each show a two dimensional projection of the resulting dark matter distribution 
at redshifts $z=31$, $z=10$ and $z=3$. The results are from a Graham simulation with 
$nc=512$ particles per dimension and a cubic lateral size of $64\ {\rm Mpc}/h$. The initial conditions 
for the $\Lambda$CDM fluctuations were generated at $z=255$ and a wake was inserted at 
$z=127$.

\begin{figure}
\centering
\includegraphics[width=.35\textwidth]{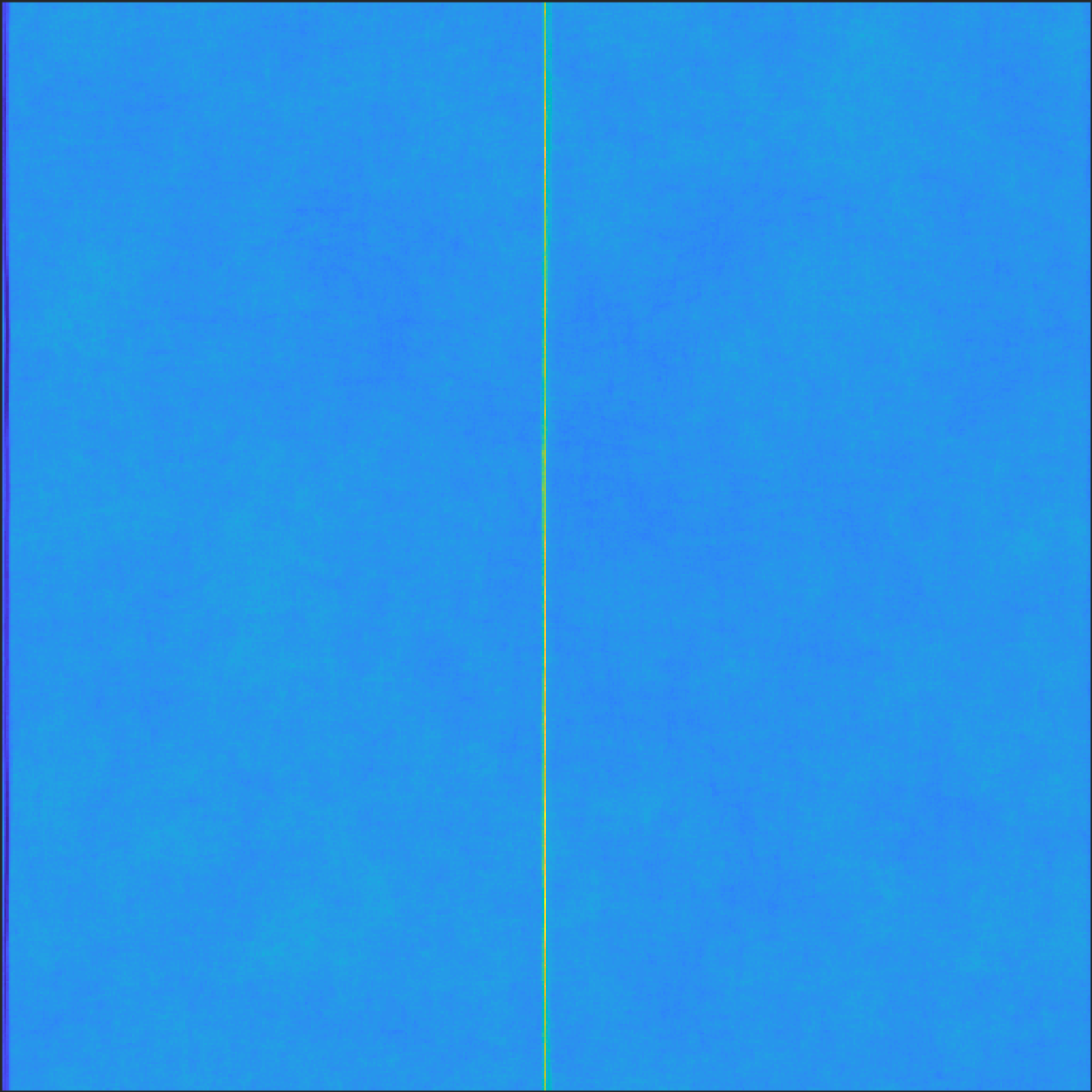}
\includegraphics[width=.35\textwidth]{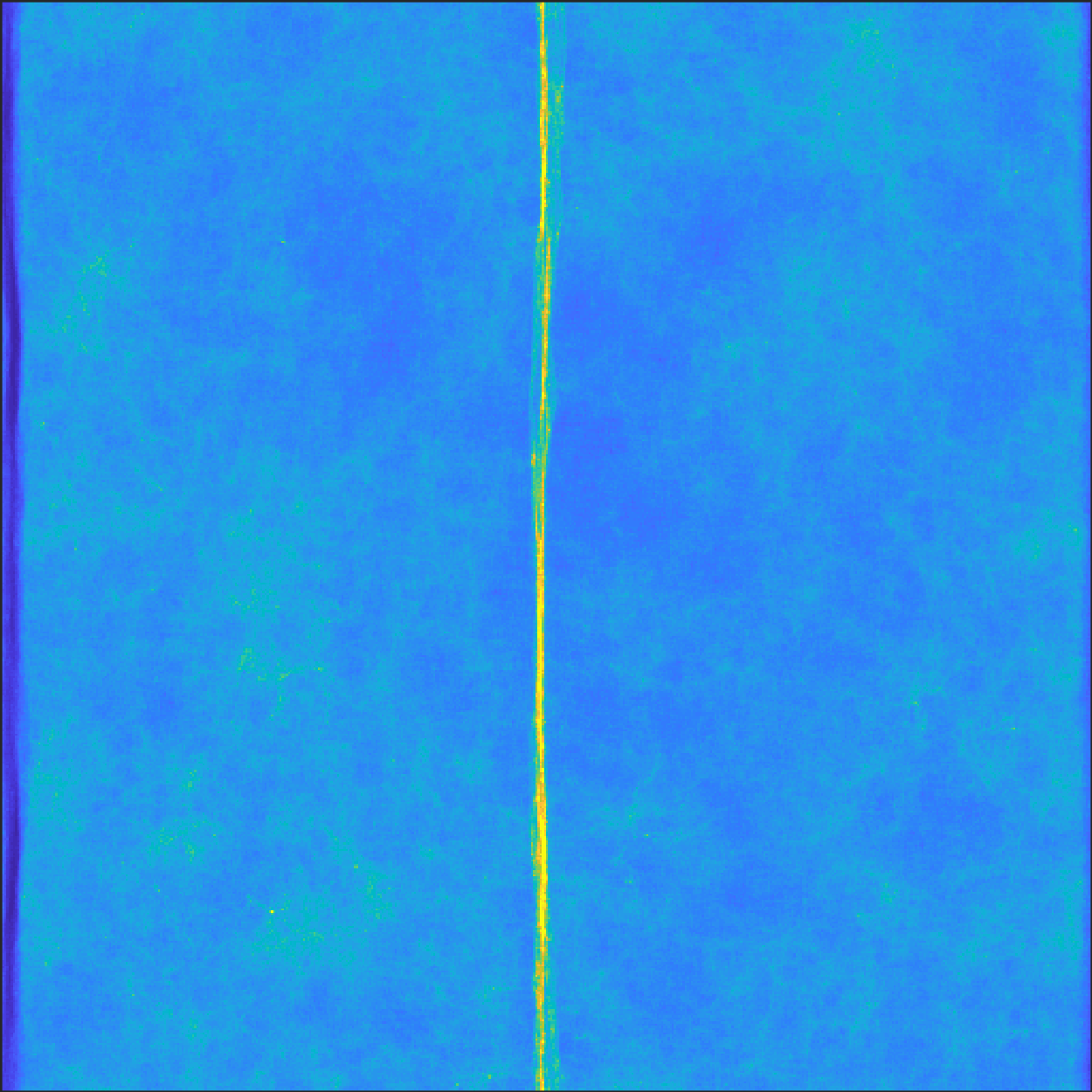}
\includegraphics[width=.35\textwidth]{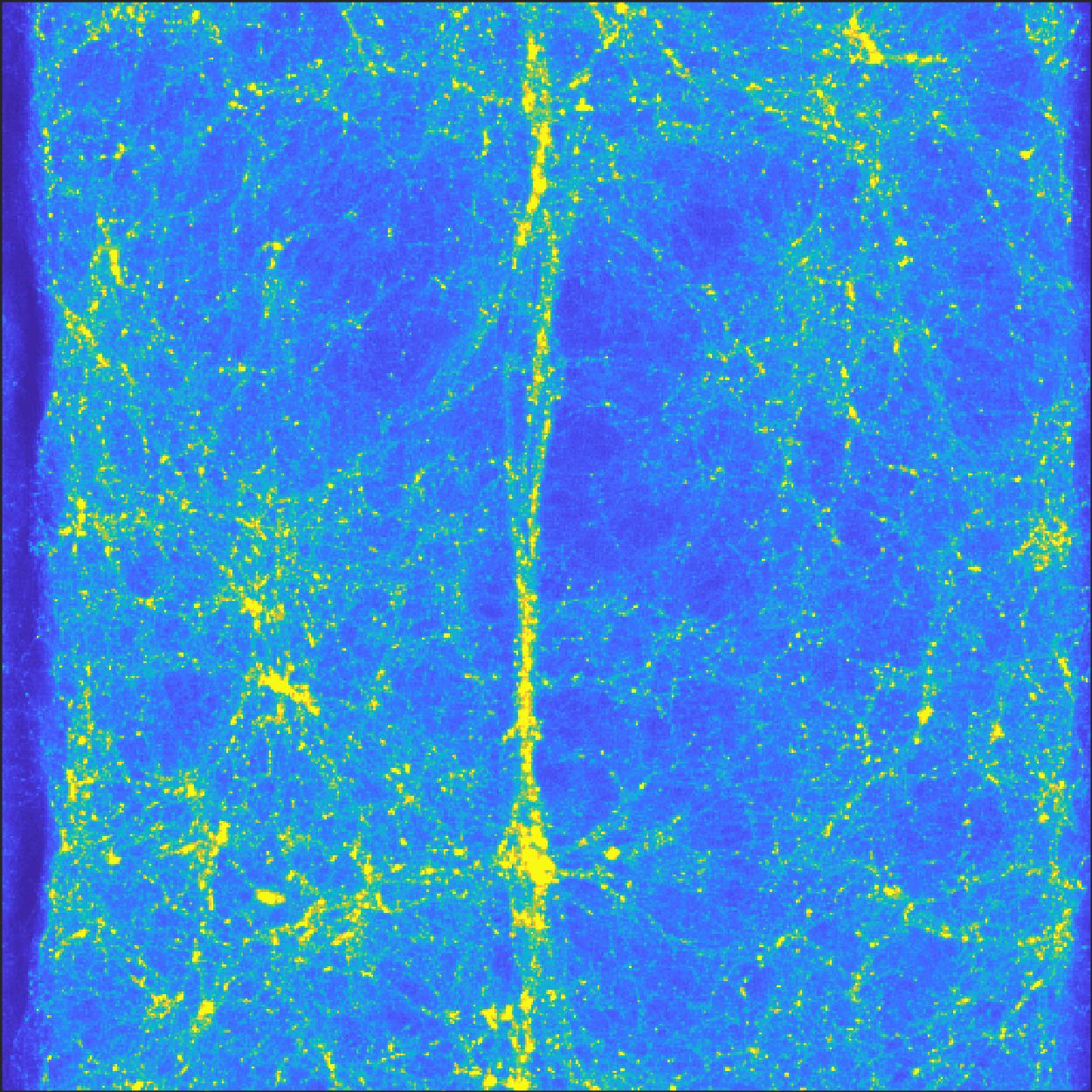}
\includegraphics[width=.4\textwidth]{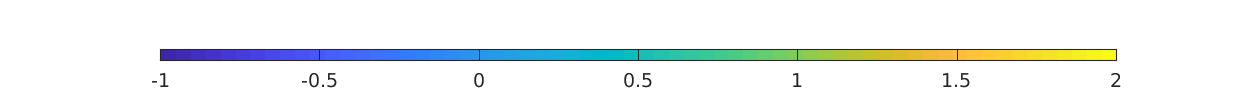}
\caption{Density contrast of the two dimensional projection of the dark matter distribution for a $G\mu=4.0\times 10^{-6}$ wake at redshift $z=31,10$ and $3$ (from top to bottom, respectively). The color bar on the bottom associates each color with the corresponding density contrast}

\end{figure}

The first figure shows the completely undistorted initial wake signal. At redshift $z = 10$ the wake
is no longer perfectly uniform, and at redshift $z = 6$ the $\Lambda$CDM fluctuations have caused
major inhomogeneities in the wake, and some small deflections. However, for this large string tension
the string fluctuations still dominate.

\subsection{Known limitations}

Our numerical modeling of a string-induced wake has multiple limitations. The first one 
concerns resolution, and results from the fact that we are not always able
to resolve the wake itself, which is increasingly thinner for lower string tensions. Ideally, 
the thickness of the wake should be at least as large as the simulation cell size,
but this is not always feasible to achieve in a cosmological setup, given the computing 
resources at our disposal. For example, supposing we would like to resolve 
a wake produced by a cosmic string with tension $G\mu = 1.0\times 10^{-7}$ at
redshift $z = 7$, the grid size 
needs to be  $0.01 {\rm Mpc}/h$, which, assuming a large simulation with 8192 cells per dimension, 
corresponds to a lateral size of $\approx 57\ {\rm Mpc}/h$. 
We circumvent this computing challenge by noting that the wake has a global impact on 
the matter field, and that we do not need to resolve the initial wake exactly to detect
its presence.

A second, less intuitive, limitation arises from the wake insertion itself: once every particle 
has been moved towards the wake, a planar region parallel to the wake is left empty at the 
boundary of the simulation box. In other words, the number of particles in the simulation is 
fixed, and the dislocation of the particles that is required to create the wake (an overdense 
region in the central plane of the simulation) produce at the same time an underdense 
region at the boundary plane. Although there is indeed a compensating underdensity
at large distances from the wake (this is required by the ``Traschen integral constraints''
\cite{Traschen} on density fluctuations in General Relativity), the fact that the void 
occurs at the boundary of our simulation box is unphysical. Note, in particular, that in 
the simulations the location of the underdensity depends on the box size, and it should
be pushed to the horizon size at wake formation \cite{Joao}. In order to preserve the 
cosmological background in the simulation, we cannot introduce new particles to fill this 
empty region, and hence we have no way to get rid of this undesired effect. 

Our approach is therefore to examine whether or not this void  affects the evolution of the 
wake. We achieve this by measuring the displacement of the particles induced by the 
presence of the wake (and the corresponding void at the boundary) and comparing 
the result with the Zel'dovich approximation formula (\ref{displacement}). Each particle in the 
simulation carries an identification number, and so it is possible to compute the position of a 
particle in a simulation without a wake and compare with the position of the same particle  
in the simulation that has an inserted wake. Since the only difference between the two 
simulations is the wake insertion, by subtracting the two positions it is possible to obtain the 
displacement of this particle induced by the wake.

\begin{figure}
\includegraphics[height=7cm]{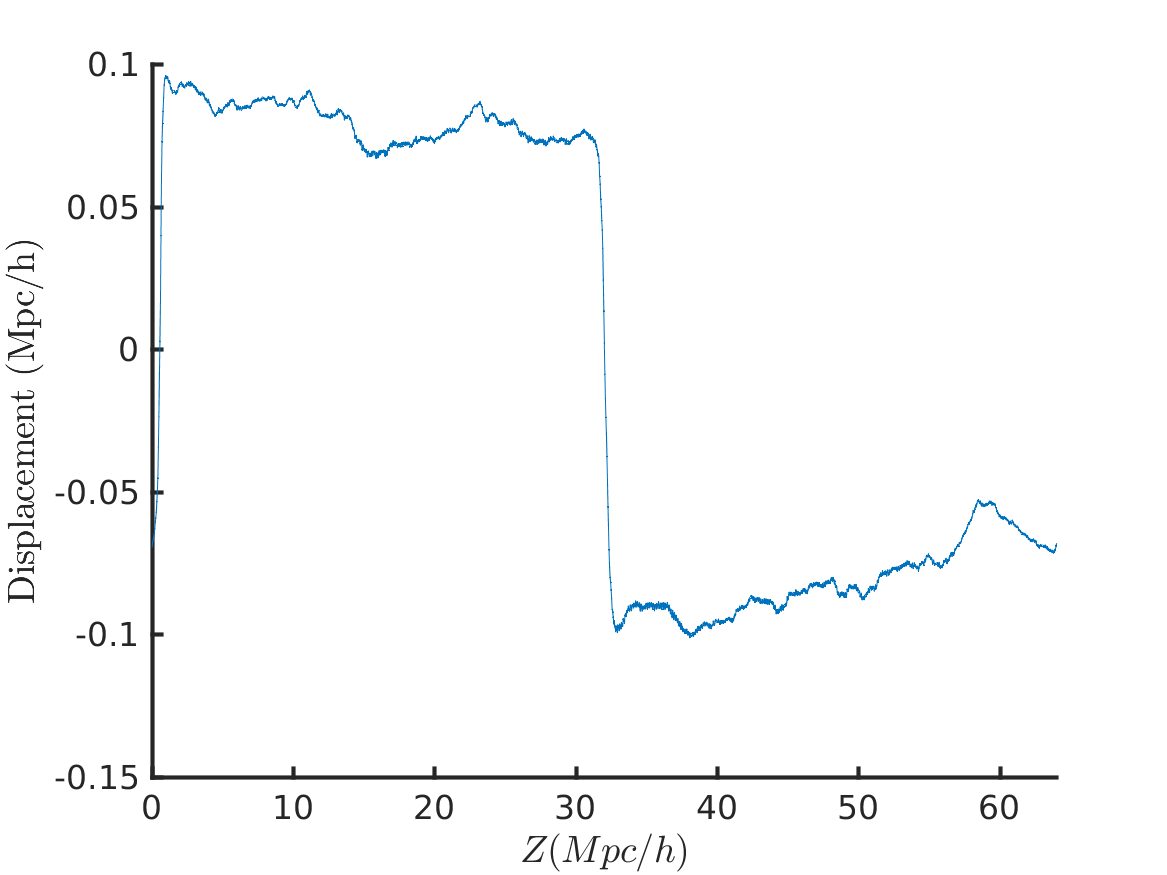}
\caption{Plot of the induced displacement due to the wake as a function of axis position 
$Z$ (horizontal axis) for redshift $z=10$ and string tension $G \mu = 8\times 10^{-7}$} 
\label{fig55}
\end{figure}

Figure \ref{fig55} shows the displacement in the direction perpendicular to the wake induced 
on the particles by the presence of the wake. A Guillimim simulation with 
$G\mu=8\times 10^{-7}$, number of cells per dimension $nc=512$ and lateral size $L=64\ {\rm Mpc}/h$ 
was used and the figure corresponds to $z=10$, with a wake inserted at $z=31$. The axis 
perpendicular to the wake was divided into bins with the same thickness as the cell size and the 
displacement associated to a given bin was computed by averaging over the displacements 
of all the particles inside it. The particles on the left receive a positive dislocation 
(towards the wake at the center) and the particles on the right side receive a negative 
displacement towards the wake, as expected. 

\begin{figure}
\includegraphics[height=7cm]{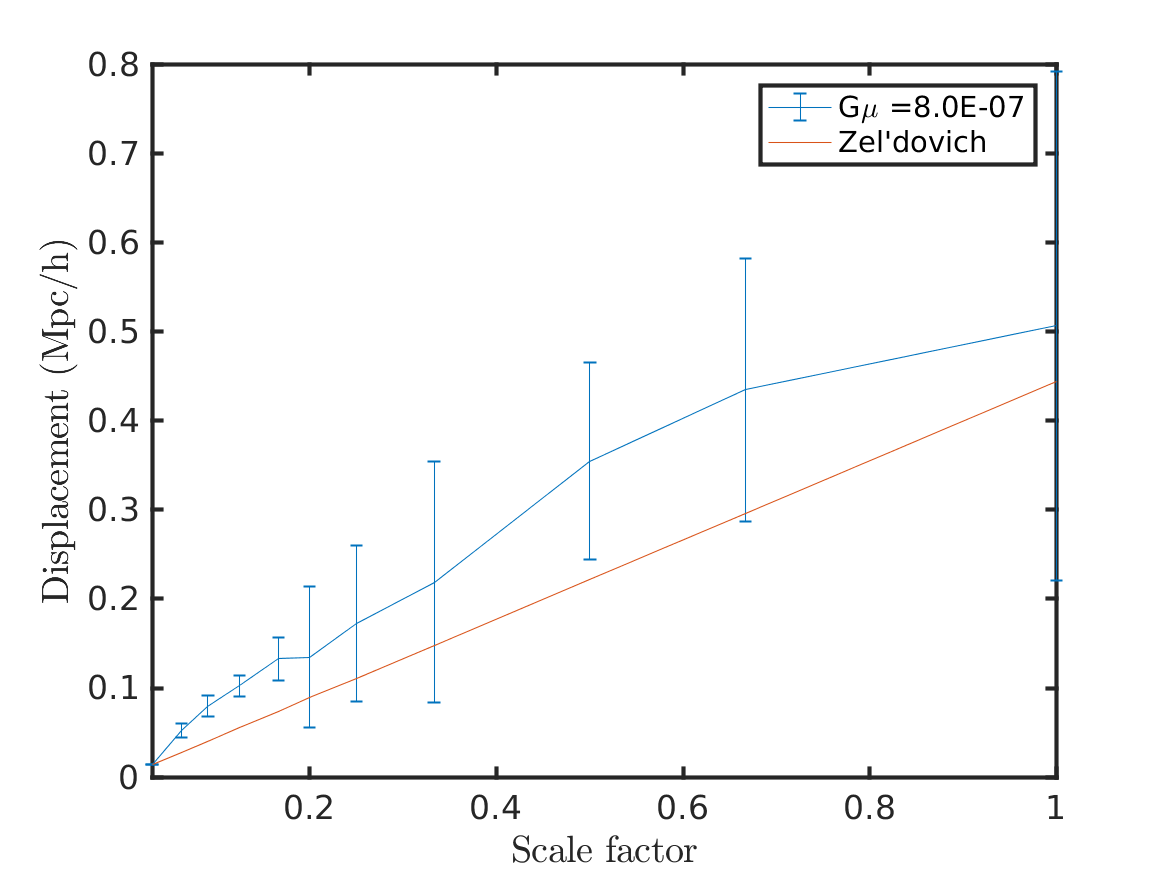}
\caption{Plot of the displacement induced by the wake (in blue) for different values of the scale factor. The expected displacement evolution from Zel'dovich approximation is shown in red.} 
\label{fig56}
\end{figure}

A number associated to the displacement associated to the above case was computed 
by considering the mean of the positive part, the absolute value of the mean of the 
negative part and taking the average of those two quantities. The error associated 
with this displacement computation was the average of the standard deviation of 
each part (positive and negative). Figure \ref{fig56} is a summary of this computation 
for different redshifts together with the expected result from the Zel'dovich approximation.

It can be seen from the results that the displacement induced by the wake in the simulation 
grows linearly in the scale factor as it should. However, in the worst case, it is about two 
times higher than the expected value.

%

\subsection{Simulations}

After these checks of the numerical code we will turn to the Guillimin ``production'' runs. We performed
$10$ simulations without wakes. From the $10$ samples, the first three were chosen for wake
insertion and evolution, adding to the dataset three samples with $G\mu = 10^{-7}$ wakes and
three with $G\mu = 8 \times 10^{-7}$ wakes. The lower value of $G\mu$ was chosen to be
just slightly below the current limit on the string tension, the second one is a larger value for
which the string effects are manifest and which can be used as a guide for the analysis.

All simulations have a grid of $1024$ cells per dimension and $512$ particles per dimension.
The volume of the simulations is $(64)^3 (h^{-1}{\rm Mpc})^3$. The initial conditions were laid down at
a redshift of $z = 63$, and the wake was inserted at redshift $z = 31$. To obtain a better
resolution of the wake at the time of wake insertion we also ran simulations where the wake
was inserted at redshift $z = 15$. A later time of wake insertion,  however, then leads to
simulations where the effects of the $\Lambda$CDM fluctuations on the wake are neglected
for a longer time. We will show that our final results do not depend sensitively on the redshift
of wake insertion. 

\begin{widetext}

\begin{figure}
\centering

\includegraphics[width=.23\textwidth]{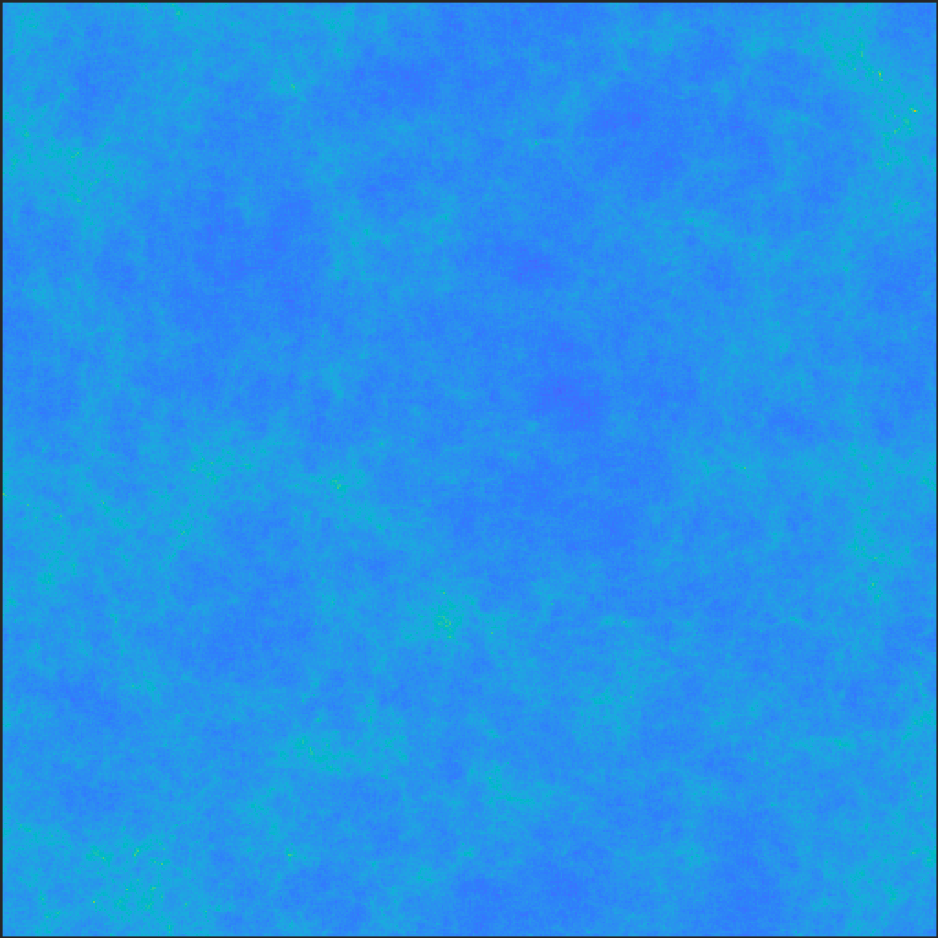}
\includegraphics[width=.23\textwidth]{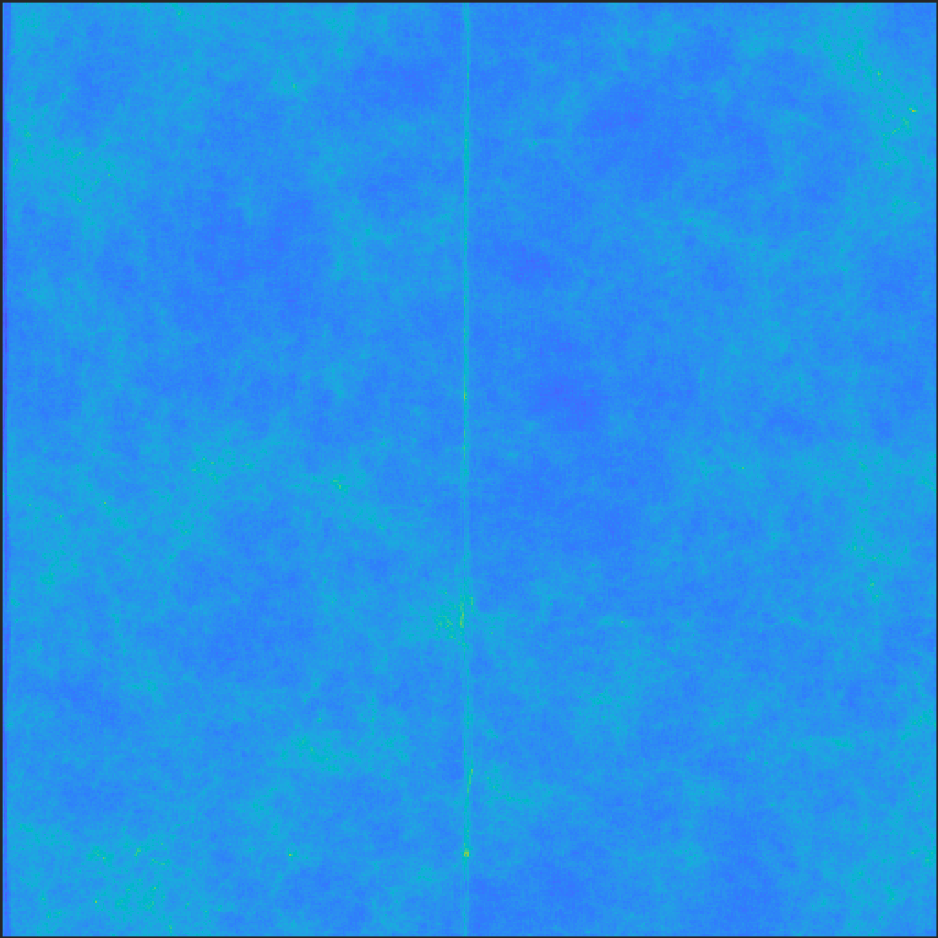}
\includegraphics[width=.23\textwidth]{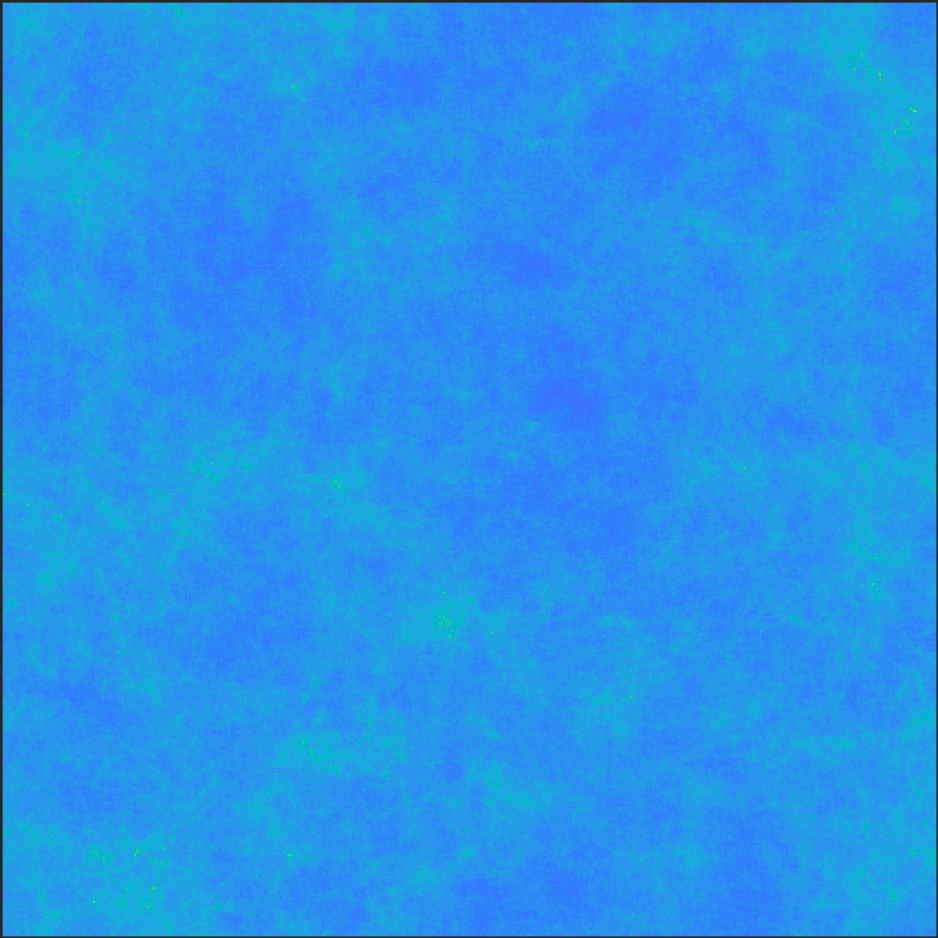}

\includegraphics[width=.23\textwidth]{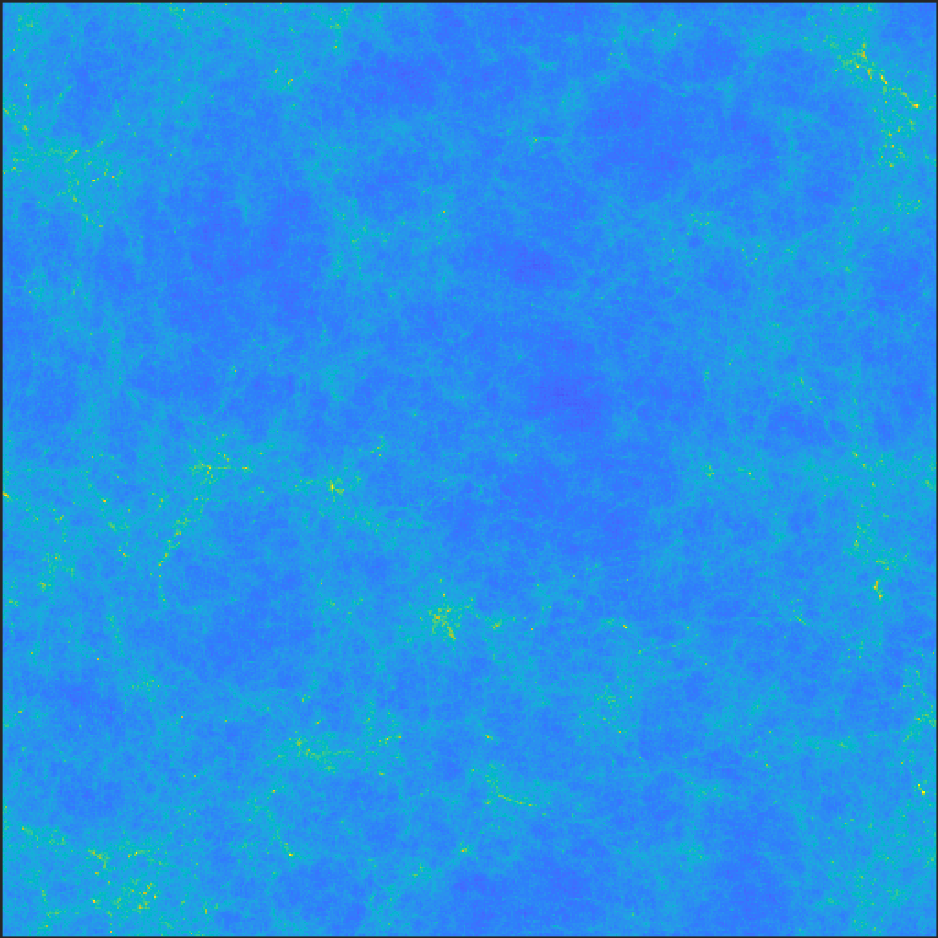}
\includegraphics[width=.23\textwidth]{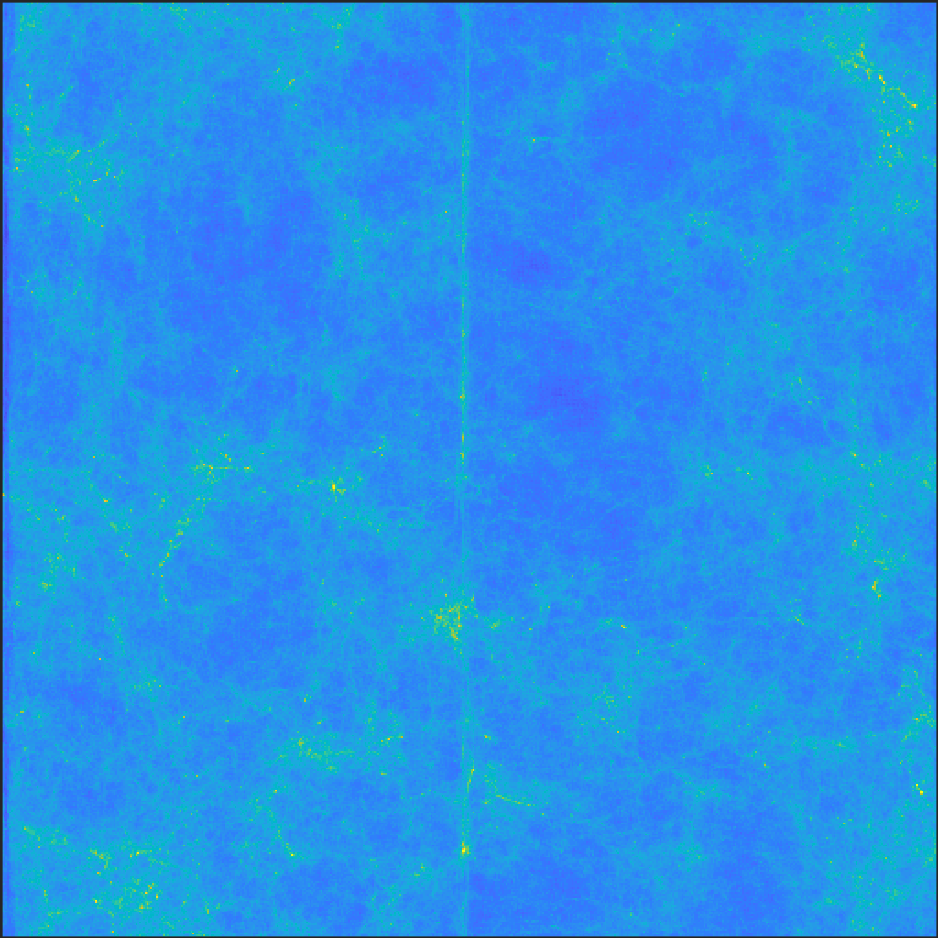}
\includegraphics[width=.23\textwidth]{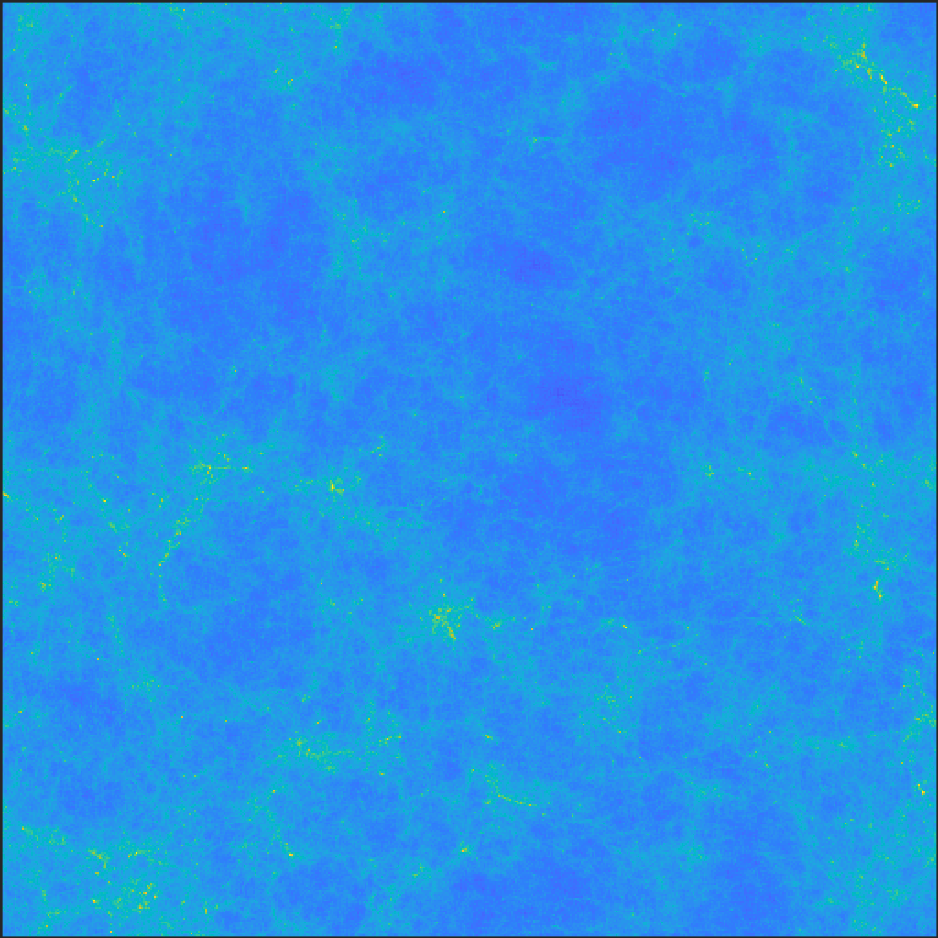}

\includegraphics[width=.23\textwidth]{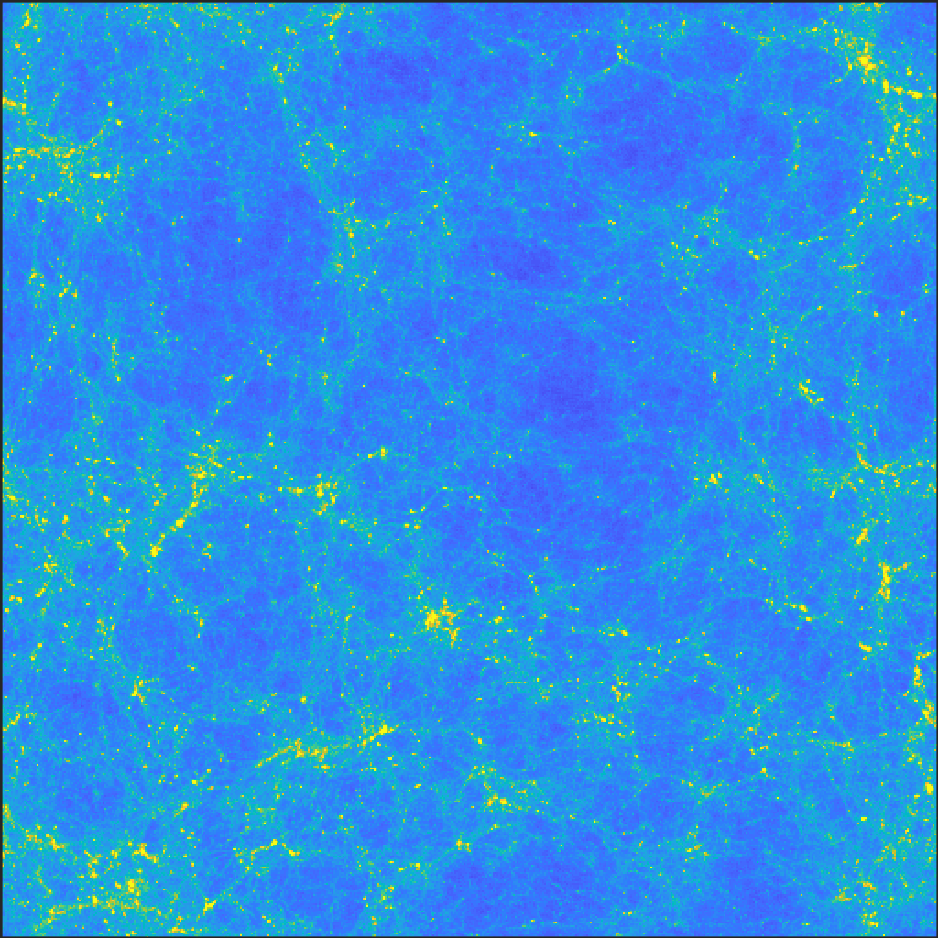}
\includegraphics[width=.23\textwidth]{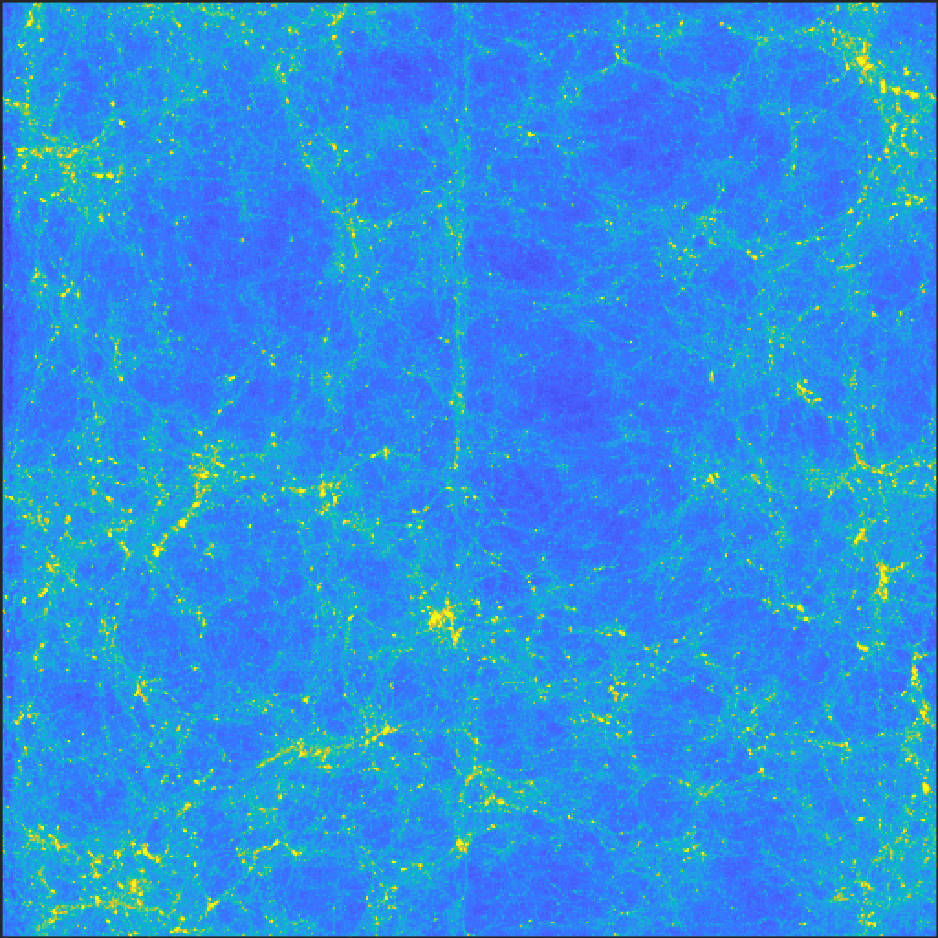}
\includegraphics[width=.23\textwidth]{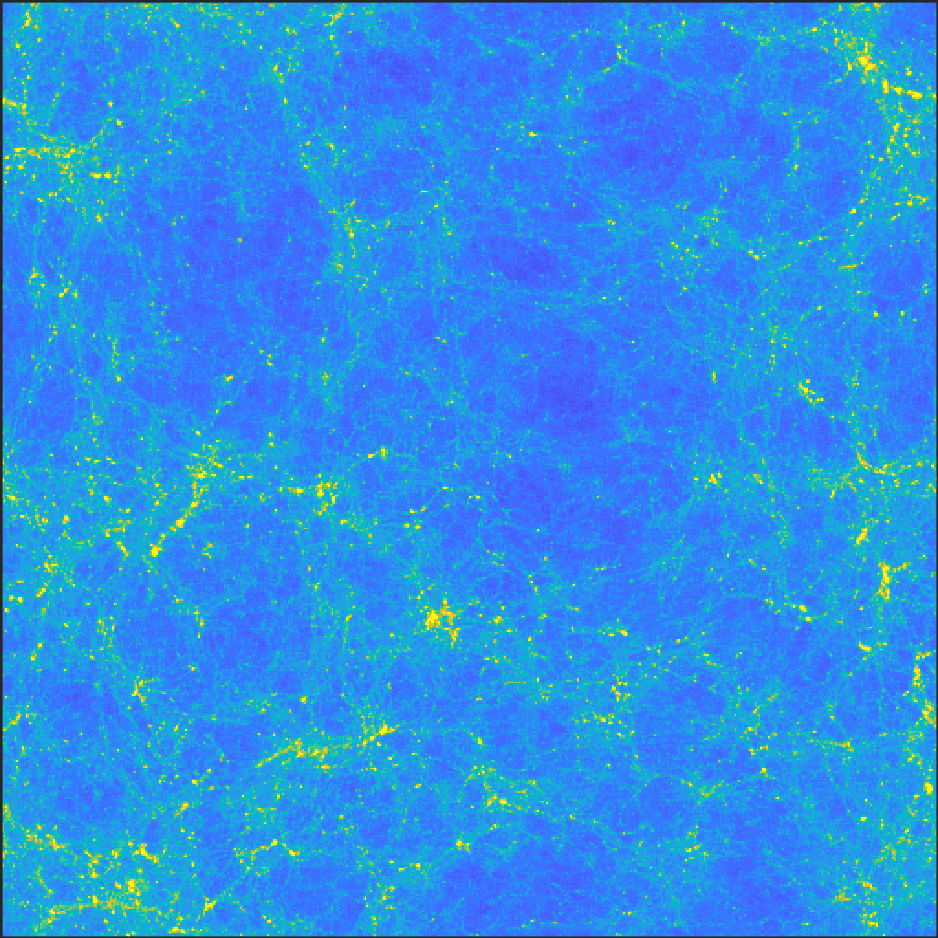}

\includegraphics[width=.23\textwidth]{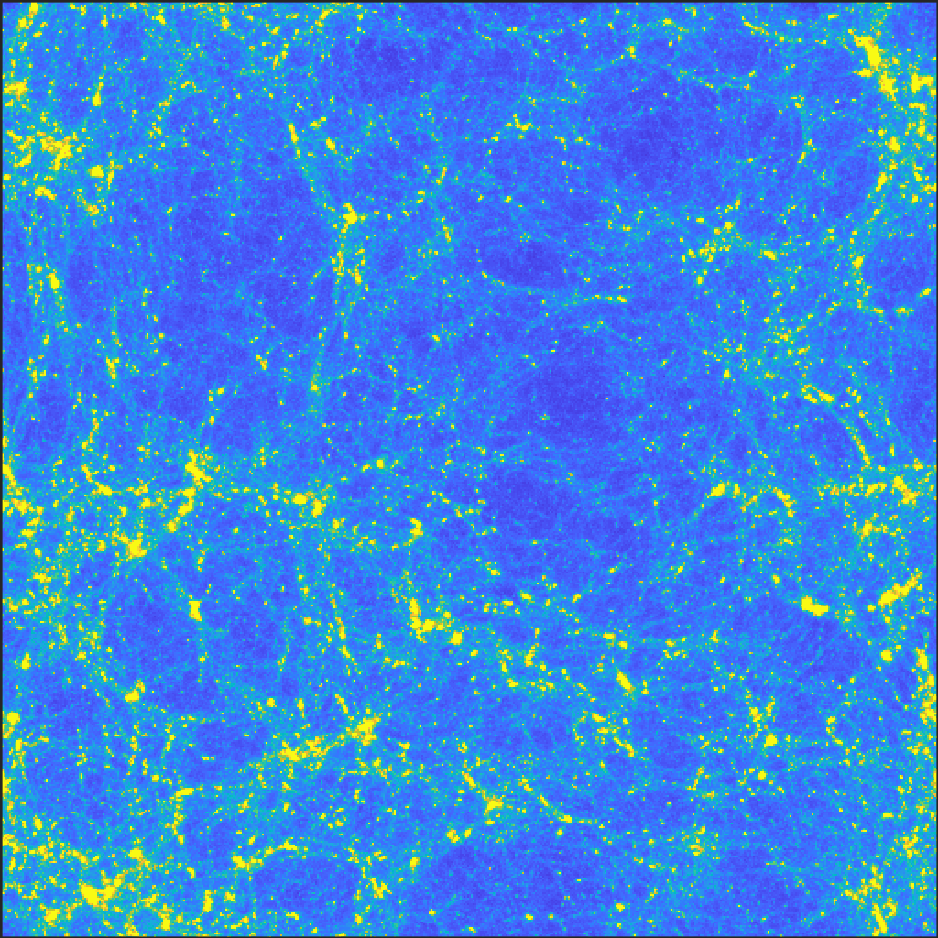}
\includegraphics[width=.23\textwidth]{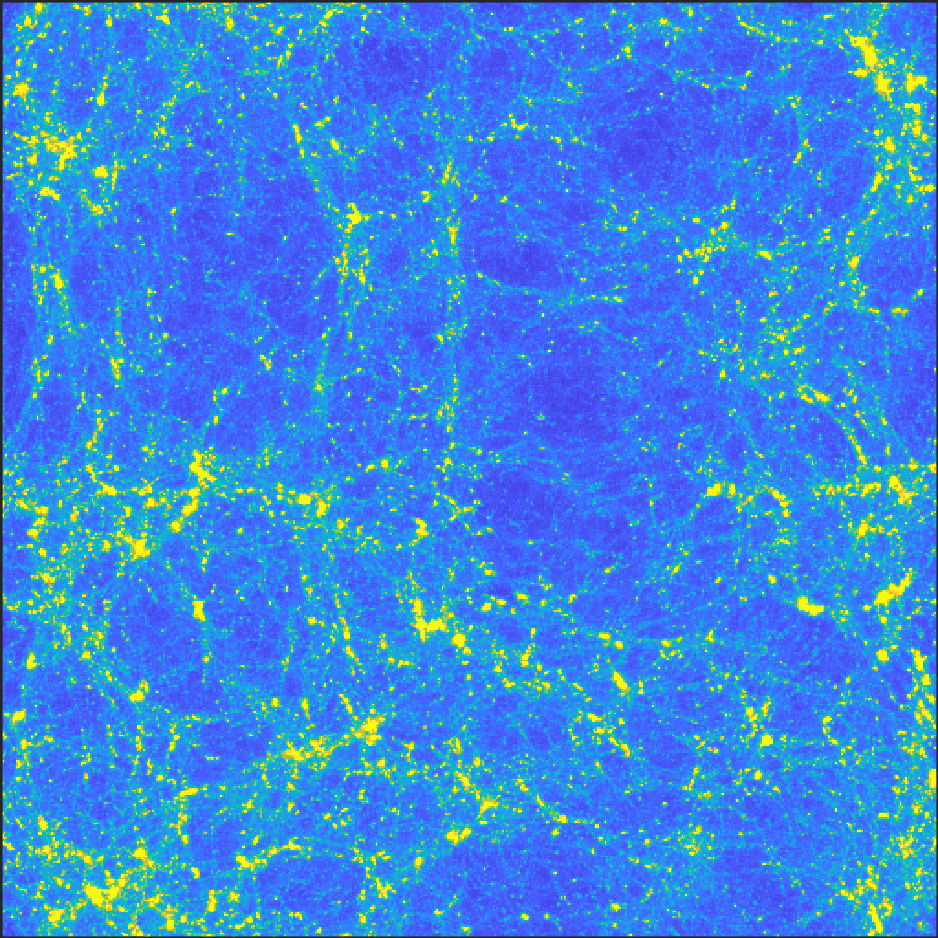}
\includegraphics[width=.23\textwidth]{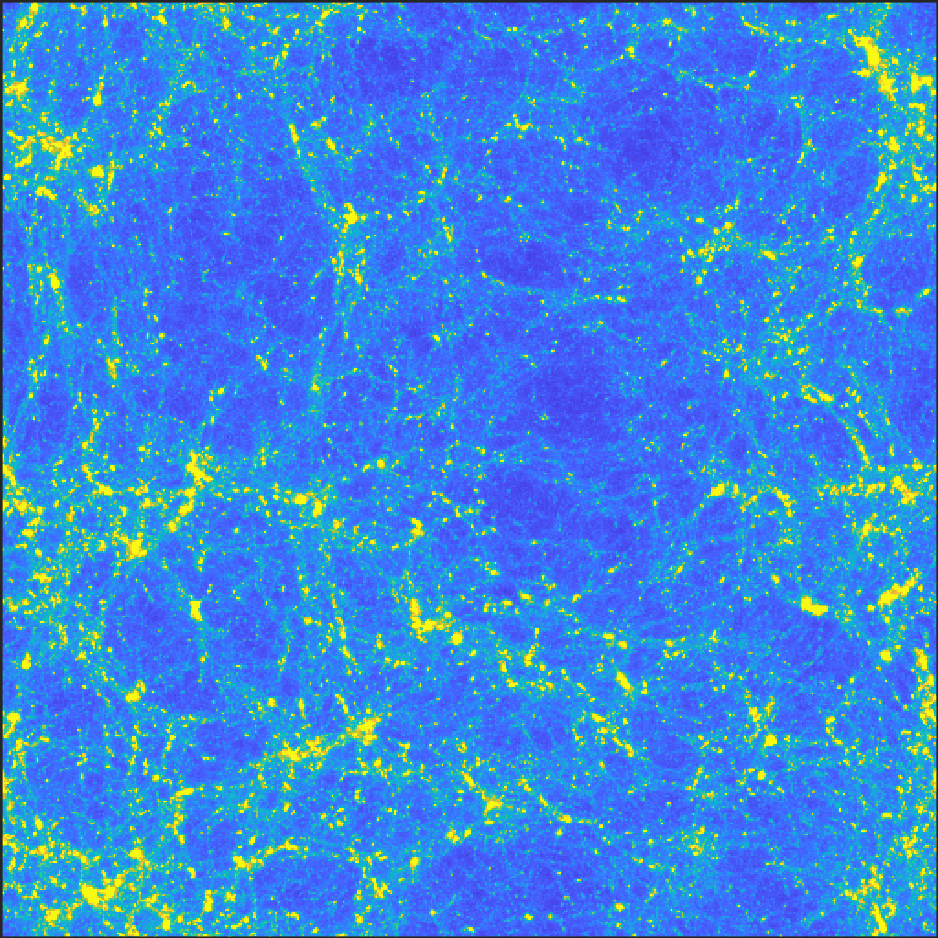}

\includegraphics[width=.23\textwidth]{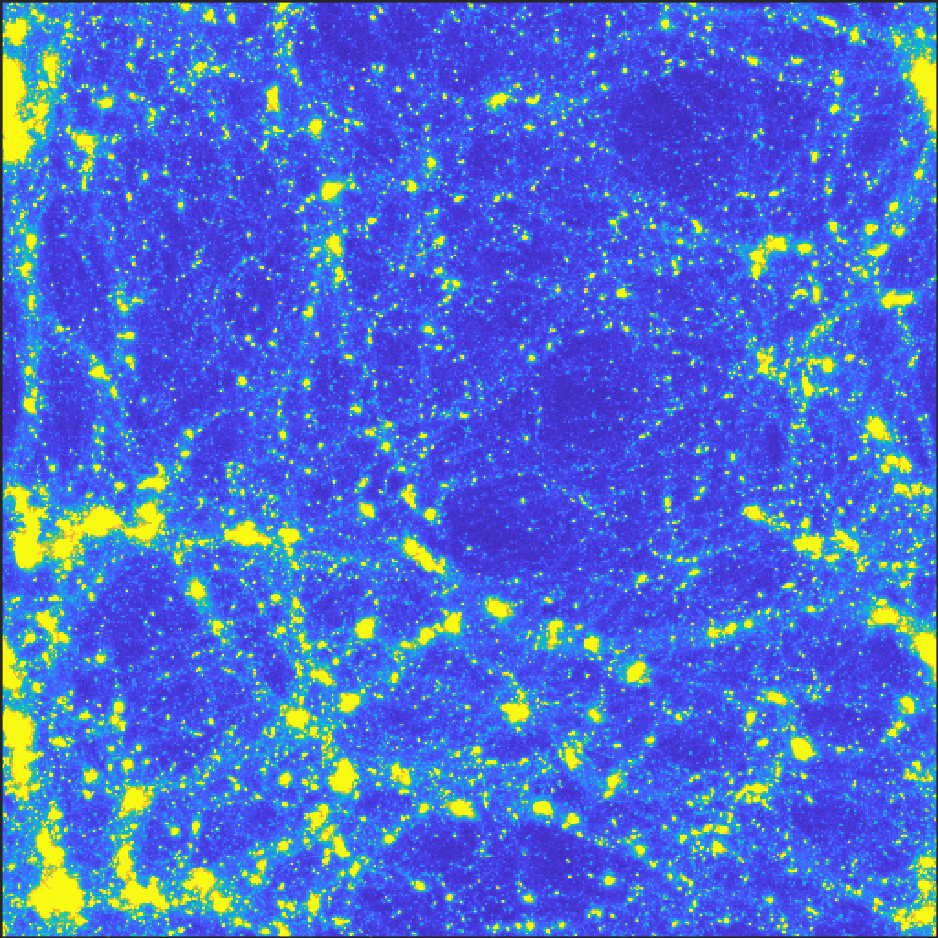}
\includegraphics[width=.23\textwidth]{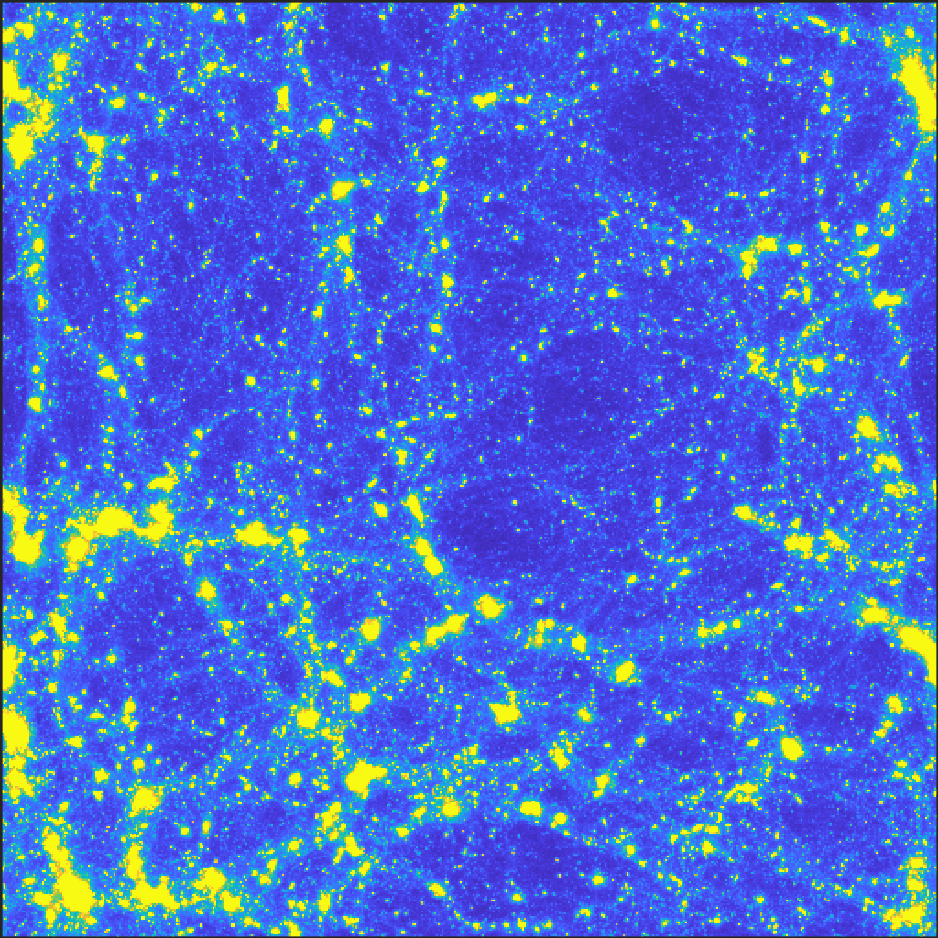}
\includegraphics[width=.23\textwidth]{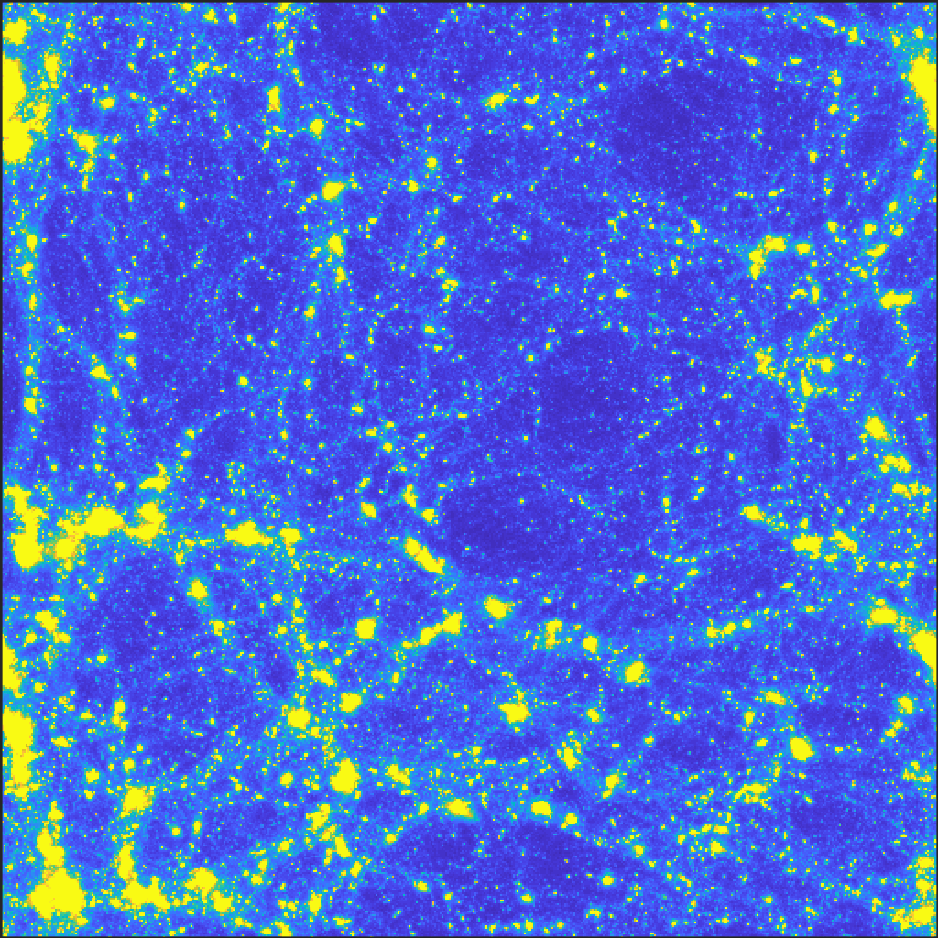}

\includegraphics[width=.8\textwidth]{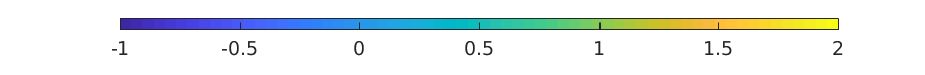}

\caption{Each panel shows the density contrast of the density field projected in a $(64 h^{-1} {\rm Mpc}/h)^2$ area. Each row depicts a redshift snapshot, chosen to be $z=10,7,4,2,0$ from top to bottom. The left column corresponds to pure $\Lambda CDM$, the middle contains a $G\mu=8\times 10^{-7}$ wake and the right column  contains a $G\mu=1\times 10^{-7}$ wake. The color bar on the bottom associates each color with the corresponding density contrast}

\label{Fig77}

\end{figure}

\end{widetext}

Figure \ref{Fig77} shows output maps of simulations at a range
of redshifts. Output map sequences of simulations without a wake, including a wake with
$G\mu = 8 \times 10^{-7}$, and a wake with $G\mu = 10^{-7}$ are shown.
The wake is placed at the center of the box along the x-axis (the horizontal
axis), and is taken to lie in the y-z plane. The y-axis is the vertical axis,
and the mass has been projected along the z-direction.

Note that the initial thickness of the wake is $5\%$ of the resolution of the simulation.
With better resolution the wake would be more clearly visible, in particular at higher
redshifts. Given the same computing power, we could increase the local resolution
at the cost of reducing the total volume, and we could study the optimal values for
the identification of the string signals. This challenge is similar to the challenge on
the observational side, where observation resolution and sky coverage need to
be balanced.
 
The leftmost column of Figure \ref{Fig77} shows the resulting mass distribution for 
redshifts $z = 10, 7, 4, 2, 0$
in a simulation without a wake, the middle column gives the corresponding output
maps for a simulation including a wake with $G\mu = 8 \times 10^{-7}$, and
with the same realization of the Gaussian noise. The wake
leads to a planar overdensity of mass which is visible by eye as a linear overdensity
along the y-axis. Until redshift $z = 4$ the wake is hardly distorted by the Gaussian
noise (it appears as an almost straight line in the plots). At redshift $z = 2$ the linear
overdensity is still clearly visible, although the Gaussian perturbations dominate
the features of the map. By redshift $z = 0$ the wake has been disrupted, although
the remnants of the linear discontinuity are still identifiable. The challenge for
a statistical analysis is to extract the wake signal at the lowest redshifts in a
quantitative way. The rightmost column of Figure \ref{Fig7} shows the corresponding output maps for a 
simulation including
a wake with $G\mu = 10^{-7}$, again with the same realization of the Gaussian
noise.  In this case, the wake is a factor of $8$ thinner
and creates primordial fluctuations which are suppressed by the same factor.
The planar overdensity due to the wake is only (and even then only extremely
weakly) identifiable at redshift $z = 10$. The challenge will be to extract this
signal in a manifest way.

\section{Statistical Analysis}

\subsection{1-d Projections}

\begin{widetext}

\begin{figure}[H]
\centering

\includegraphics[width=.32\textwidth]{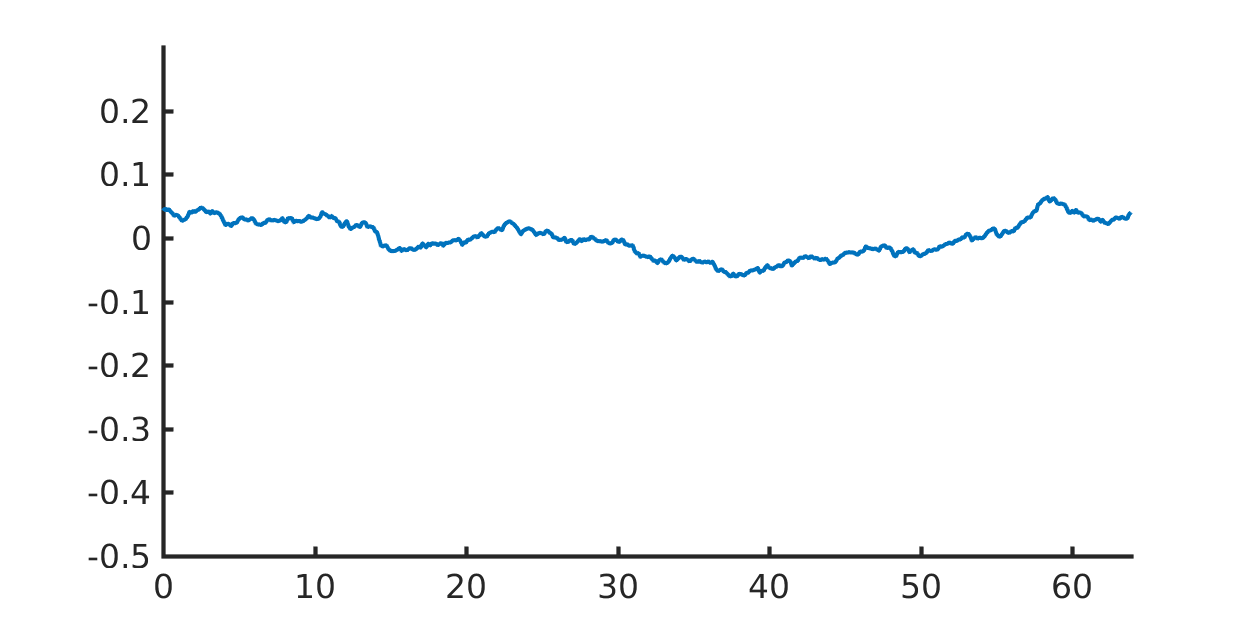}
\includegraphics[width=.32\textwidth]{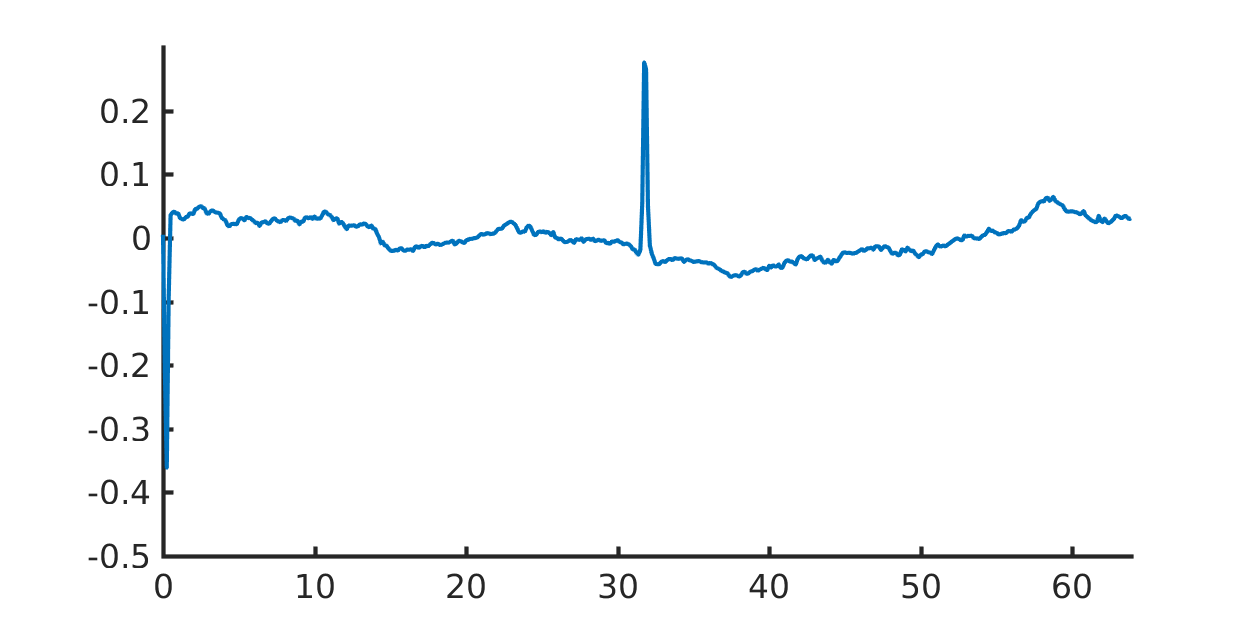}
\includegraphics[width=.32\textwidth]{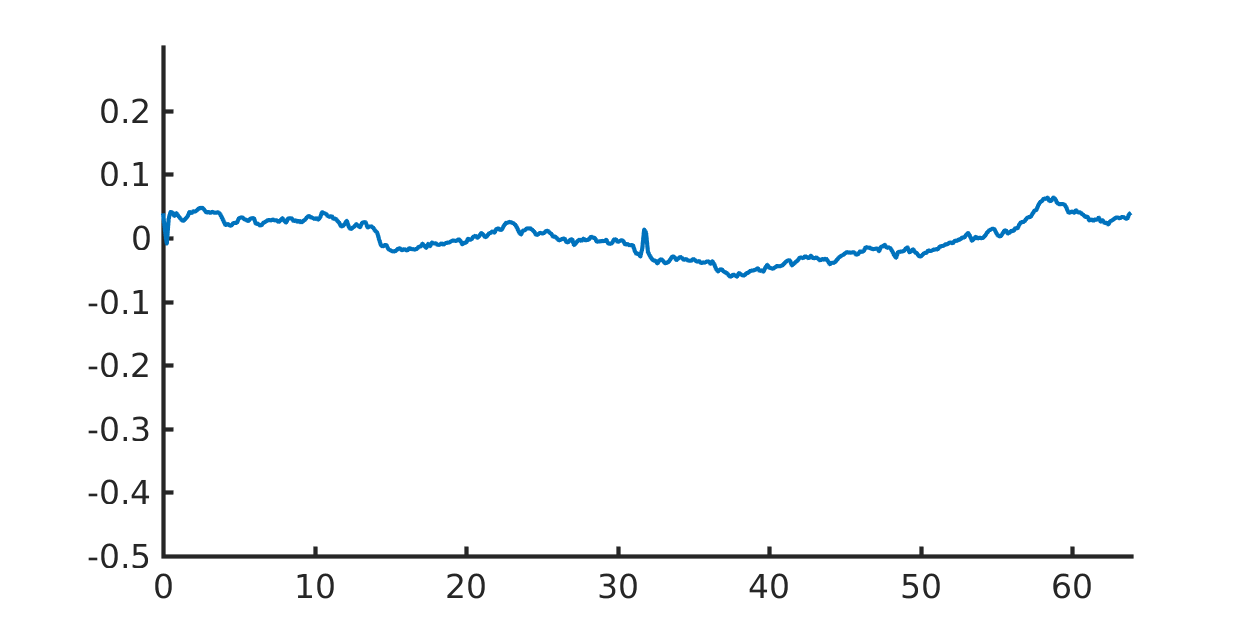}

\includegraphics[width=.32\textwidth]{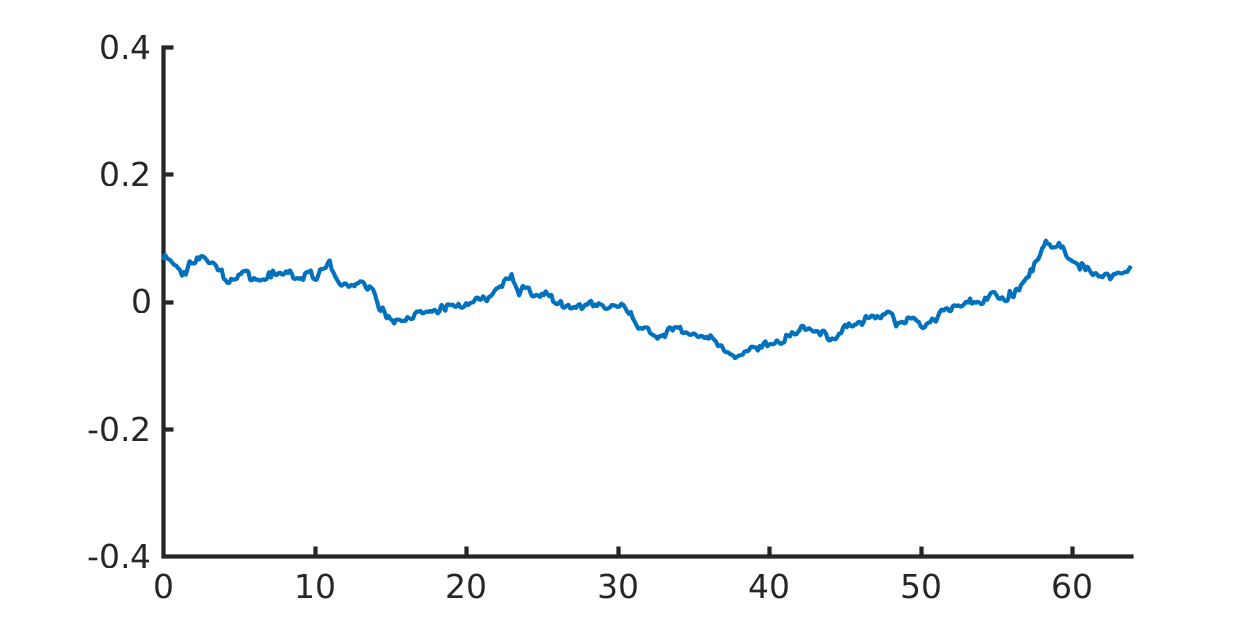}
\includegraphics[width=.32\textwidth]{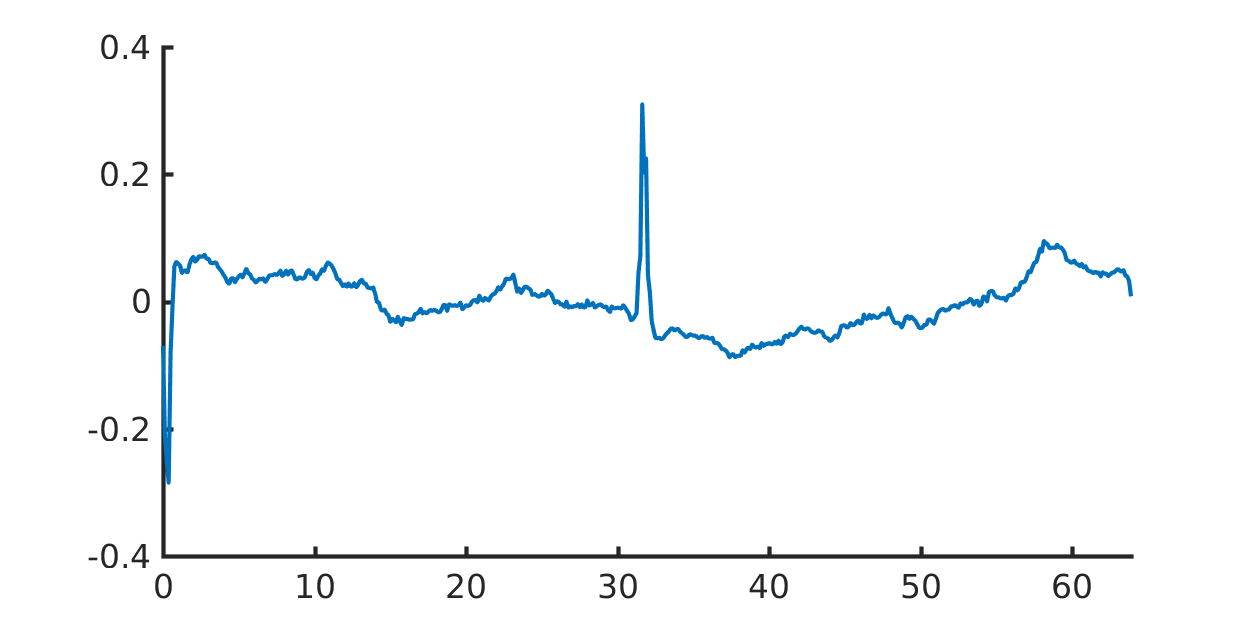}
\includegraphics[width=.32\textwidth]{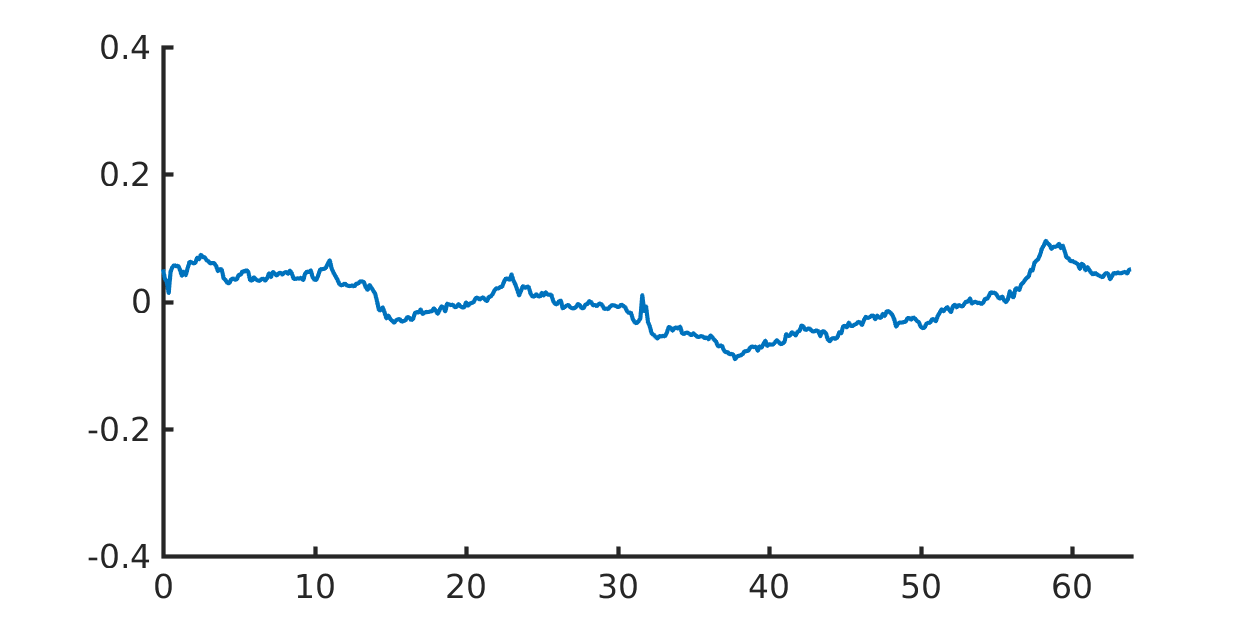}

\includegraphics[width=.32\textwidth]{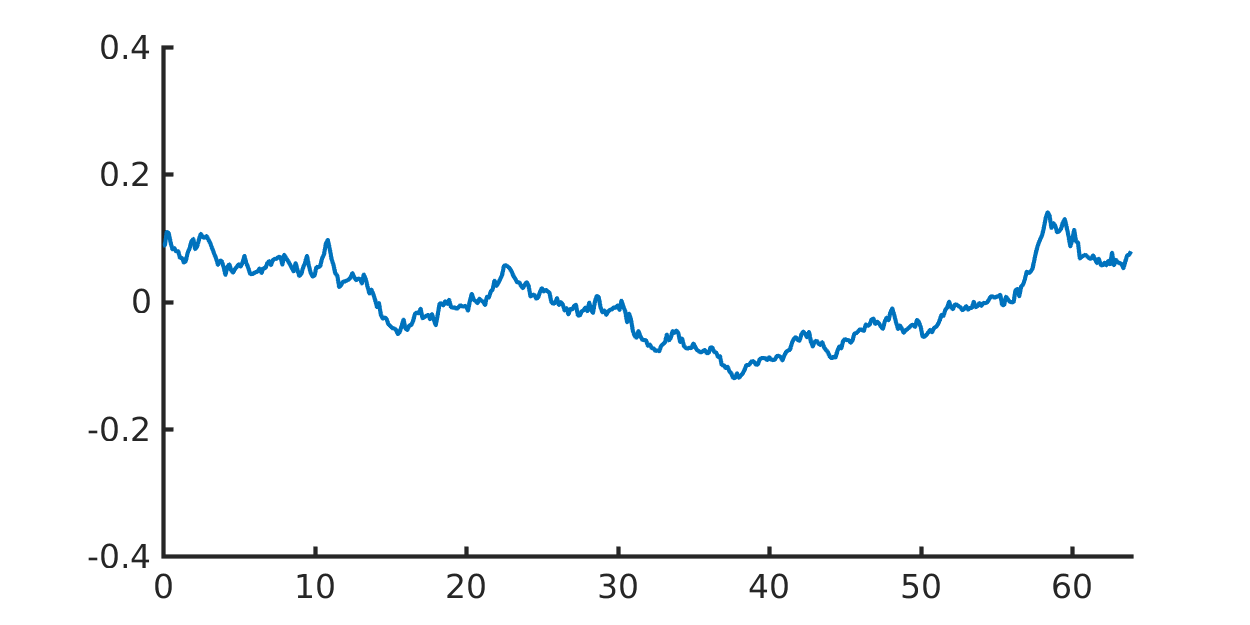}
\includegraphics[width=.32\textwidth]{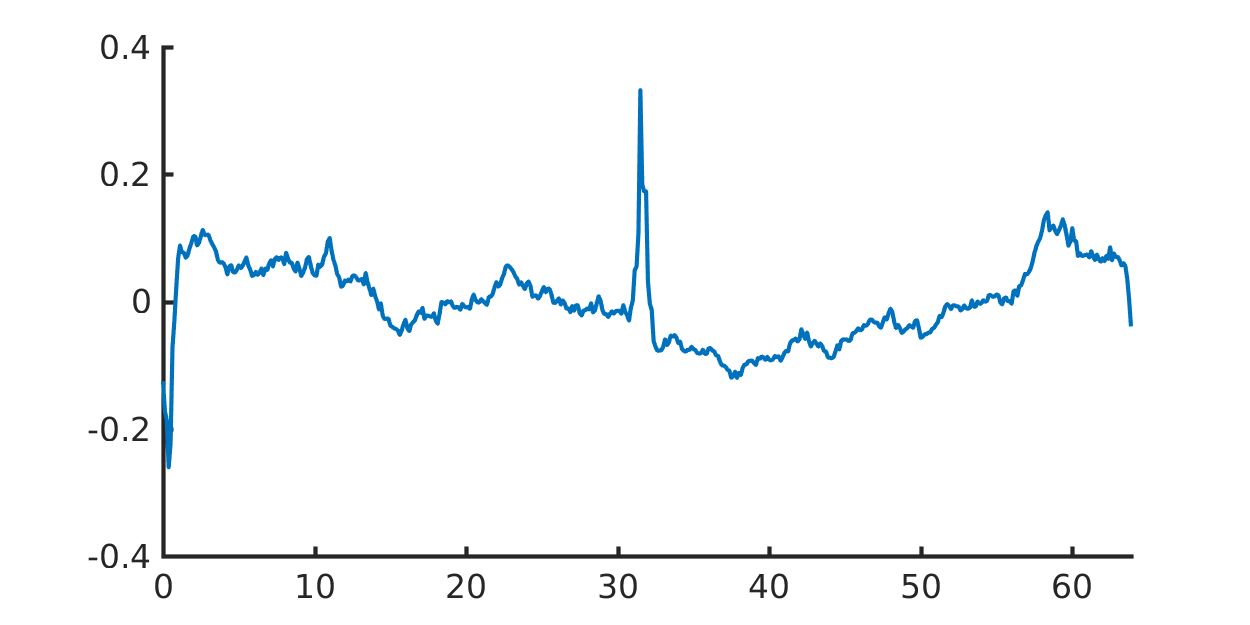}
\includegraphics[width=.32\textwidth]{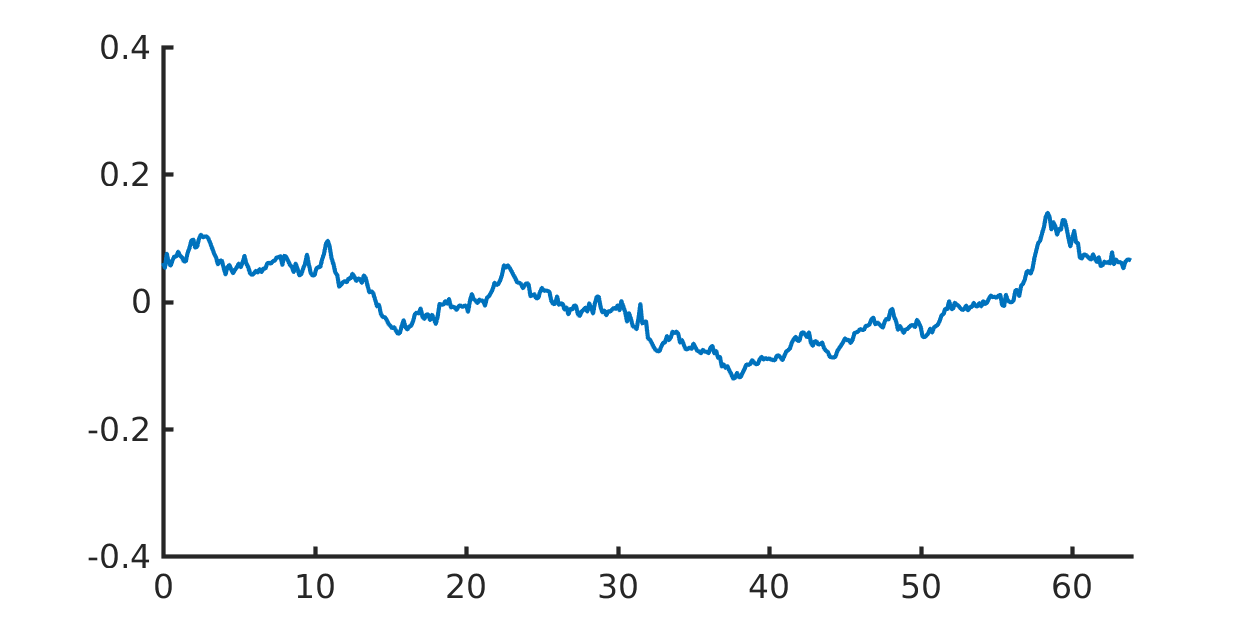}

\includegraphics[width=.32\textwidth]{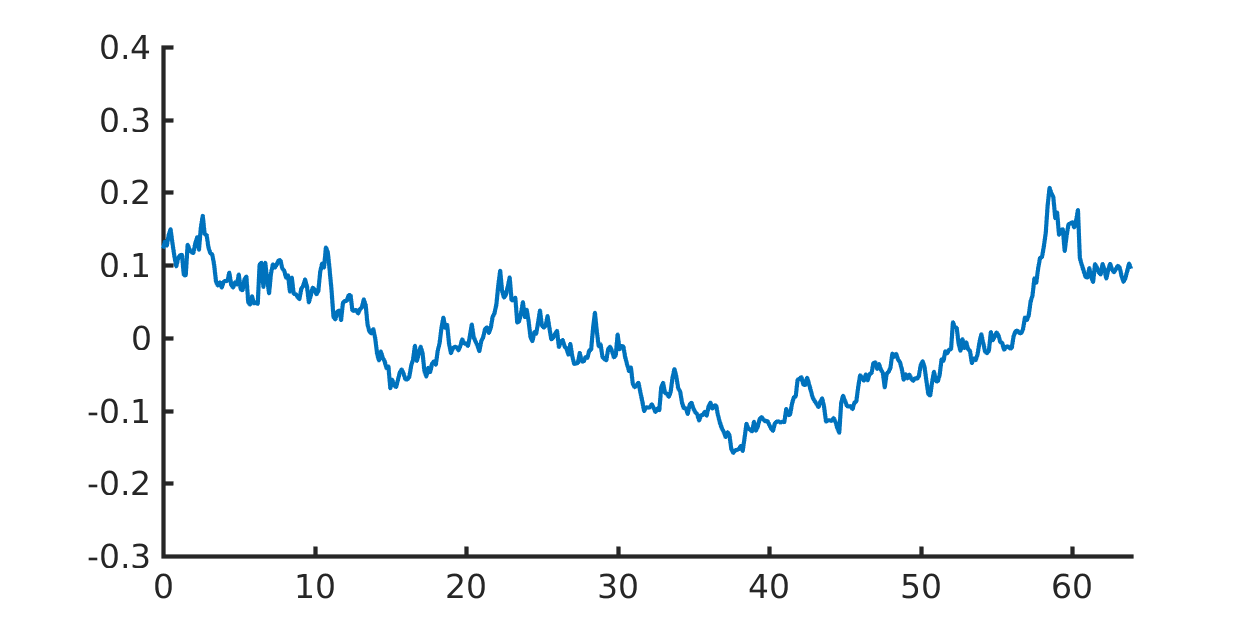}
\includegraphics[width=.32\textwidth]{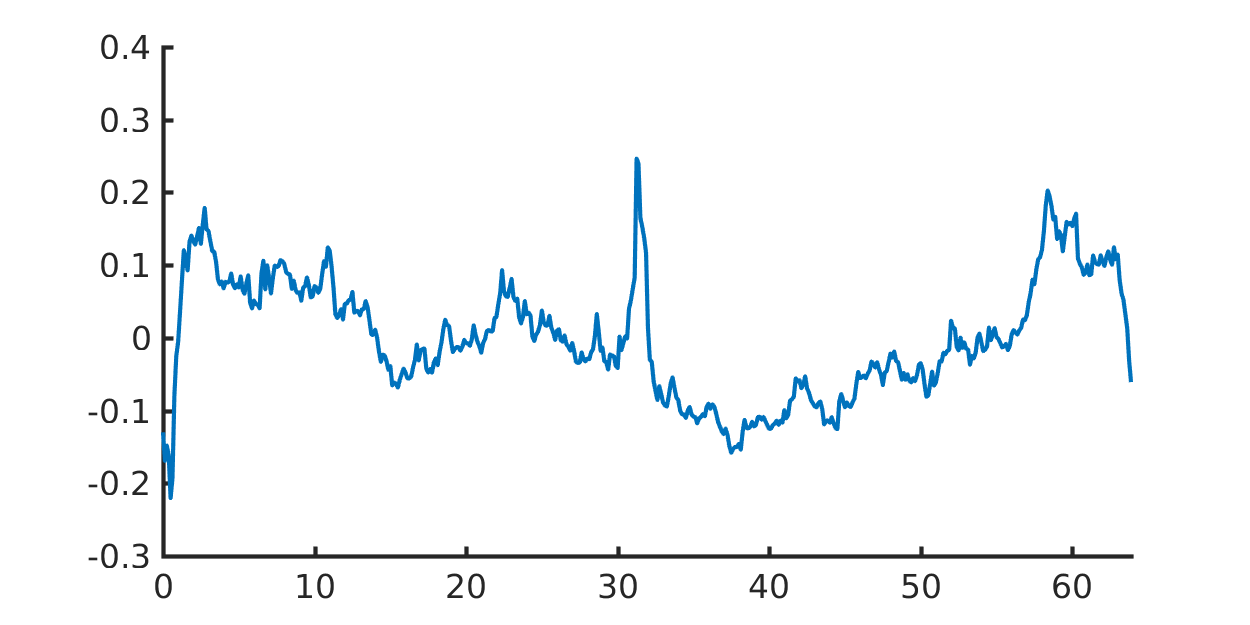}
\includegraphics[width=.32\textwidth]{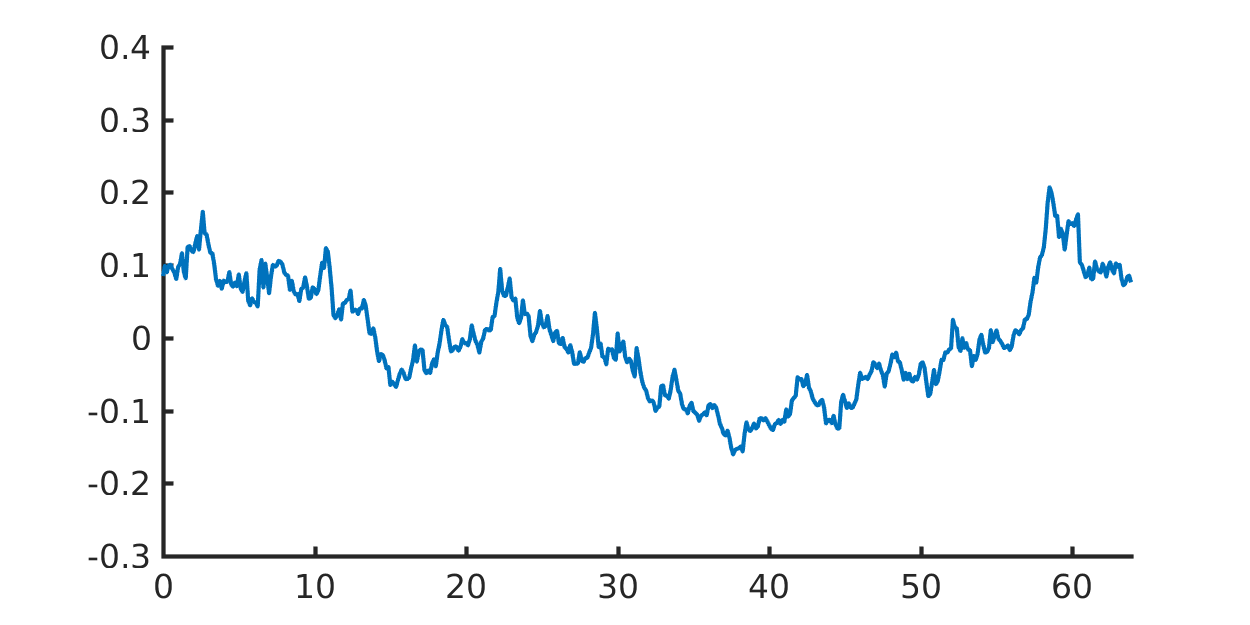}

\includegraphics[width=.32\textwidth]{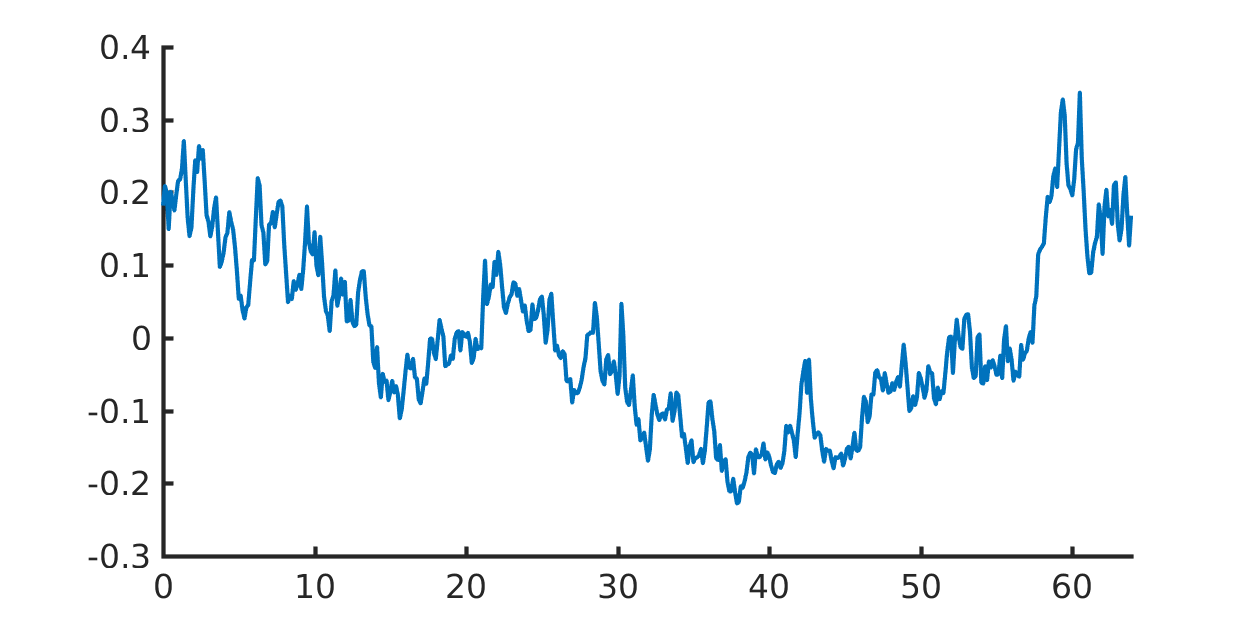}
\includegraphics[width=.32\textwidth]{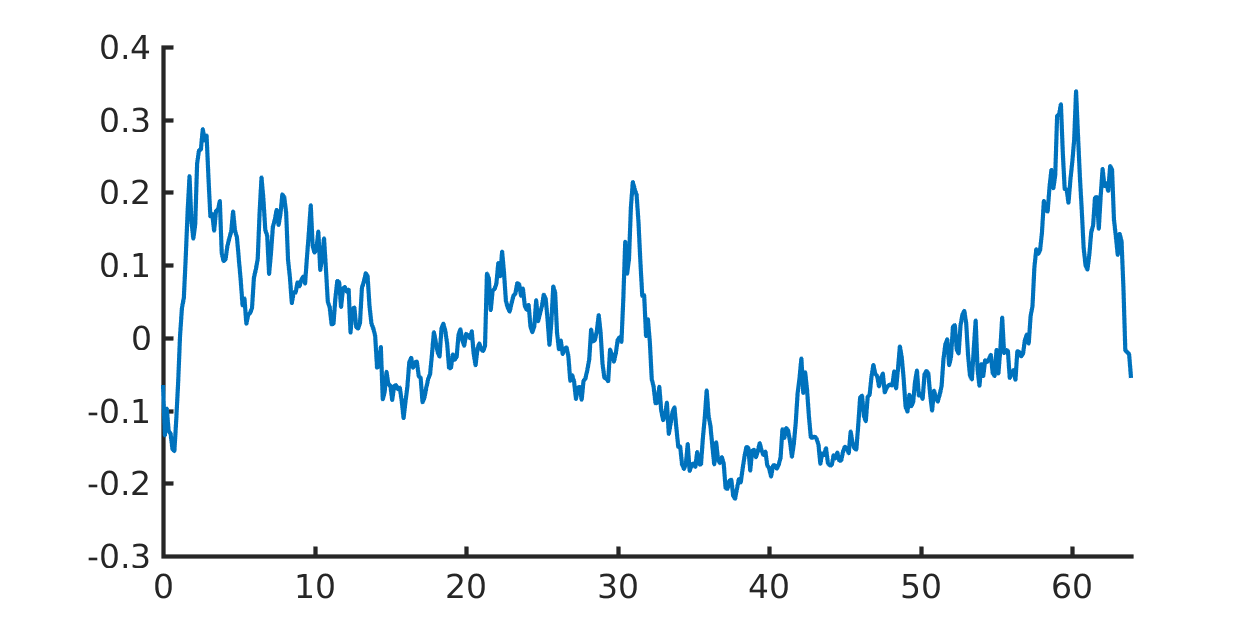}
\includegraphics[width=.32\textwidth]{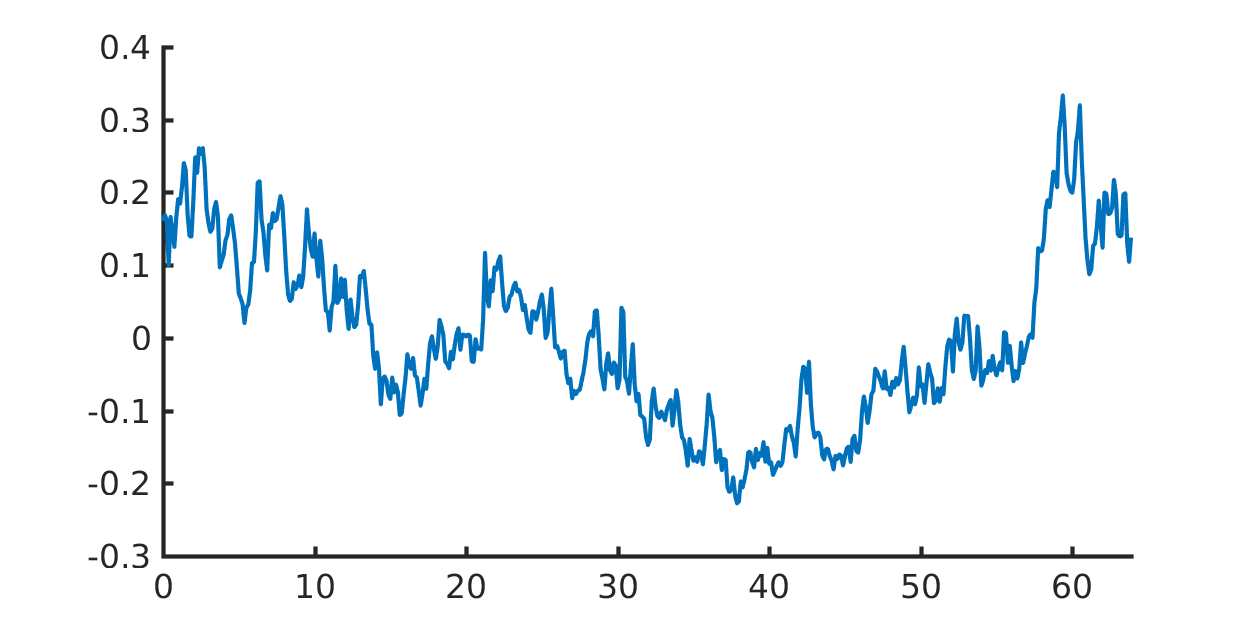}

\includegraphics[width=.32\textwidth]{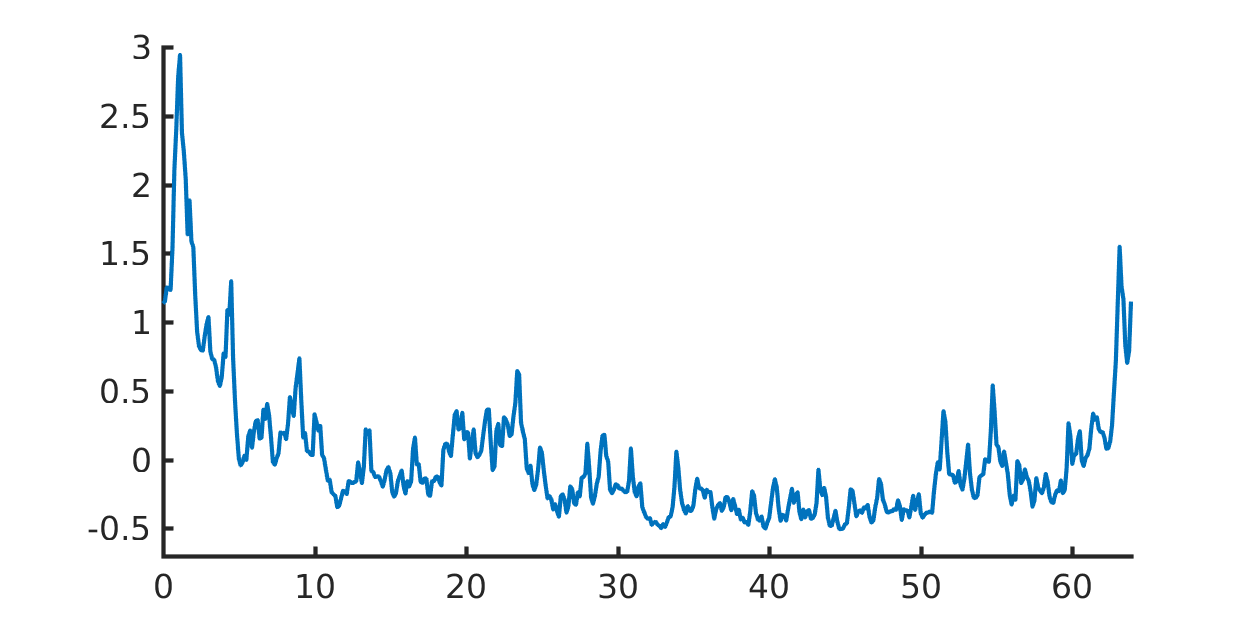}
\includegraphics[width=.32\textwidth]{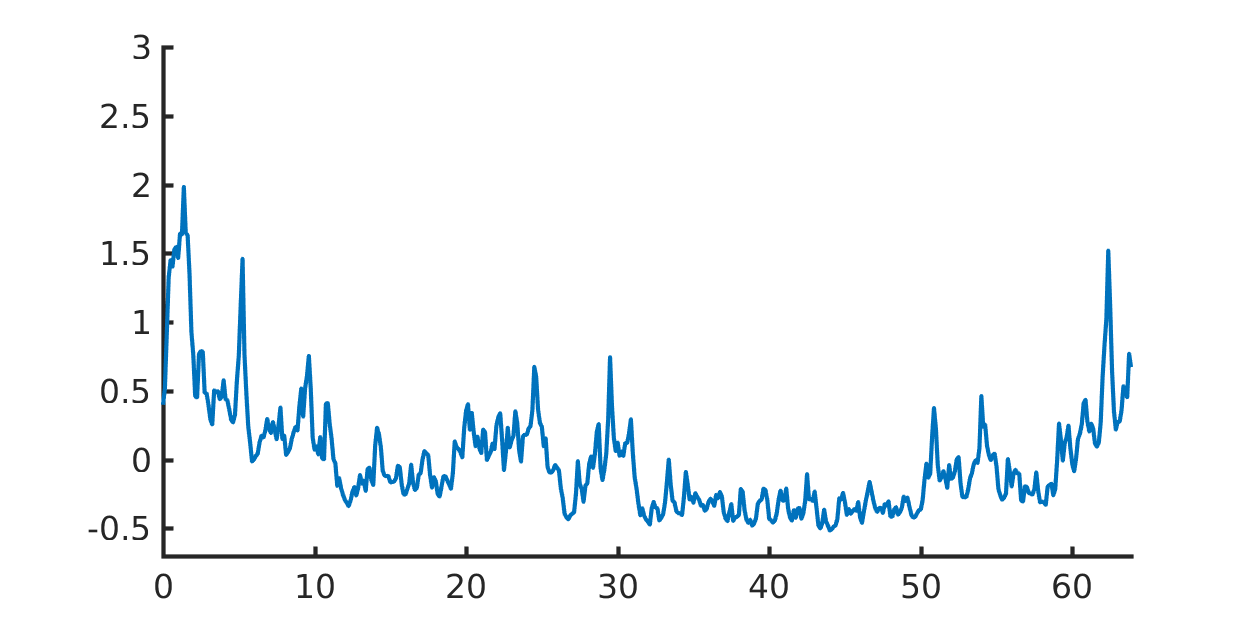}
\includegraphics[width=.32\textwidth]{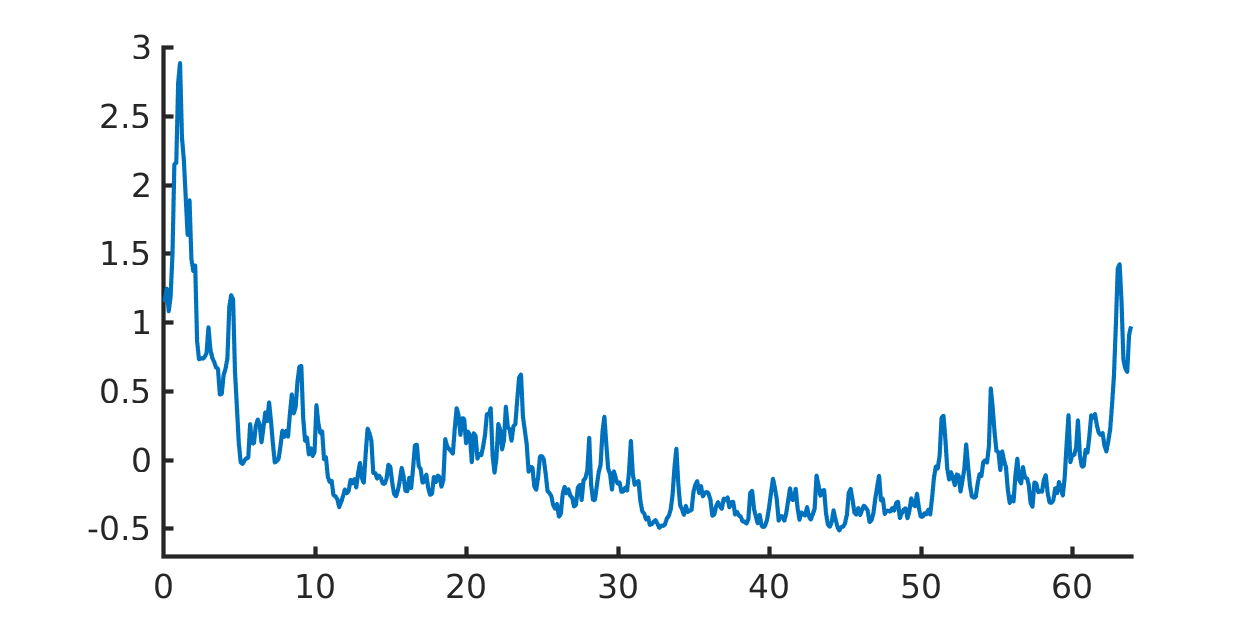}

\caption{Each panel shows the density contrast of the density field projected in a $(64\ {\rm Mpc}/h)$ axis perpendicular to the wake, when there is one. Each row depict a redshift snapshot, chosen to be $z=15,10,7,5,3,0$ from top to bottom. The left column corresponds to pure $\Lambda CDM$, the middle contains a $G\mu=8\times 10^{-7}$ wake and the right column  contains a $G\mu=1\times 10^{-7}$ wake.}

\label{Fig8}

\end{figure}

\end{widetext}

Our first step in the statistical analysis of the output maps
is to consider a further projection of the density, namely
a projection onto the direction perpendicular to the wake.
Figure \ref{Fig8}  shows the resulting distributions for a selection
of redshifts (decreasing from top to bottom) for
a simulation without a wake
(left column), including an added wake with $G\mu = 8 \times 10^{-7}$
(middle column) and $G\mu = 10^{-7}$ (right column), in both cases
with the same realization of the Gaussian noise. The vertical
axis shows the relative density contrast, the horizontal
axis is the coordinate perpendicular to the wake. The wake
corresponds to the peak located at distance $d_z \simeq 32 h^{-1} {\rm Mpc}$
At redshifts $z = 15$ and $z = 10$ 
the wake can be clearly identified by eye at this redshift even for
$G\mu = 10^{-7}$.

As the wake gets disrupted by the Gaussian noise, the
wake signal gets harder to identify at lower redshifts. For
$G\mu = 10^{-7}$ the signal can be seen at redshift
$z = 10$, but it has disappeared by $z = 7$,
while for $G\mu = 8 \times 10^{-7}$ the signal is still present at
$z = 3$, but no longer at $z = 0$.

Wakes are very thin compared to the scale where the power
spectrum of the Gaussian density perturbations peaks. This
is particularly true for lower values of the string tension. Hence, a
promising method of rendering the wake signal more visible is
to perform a wavelet transform of the 1-d projection plots.

\begin{figure}

\includegraphics[width=.5\textwidth]{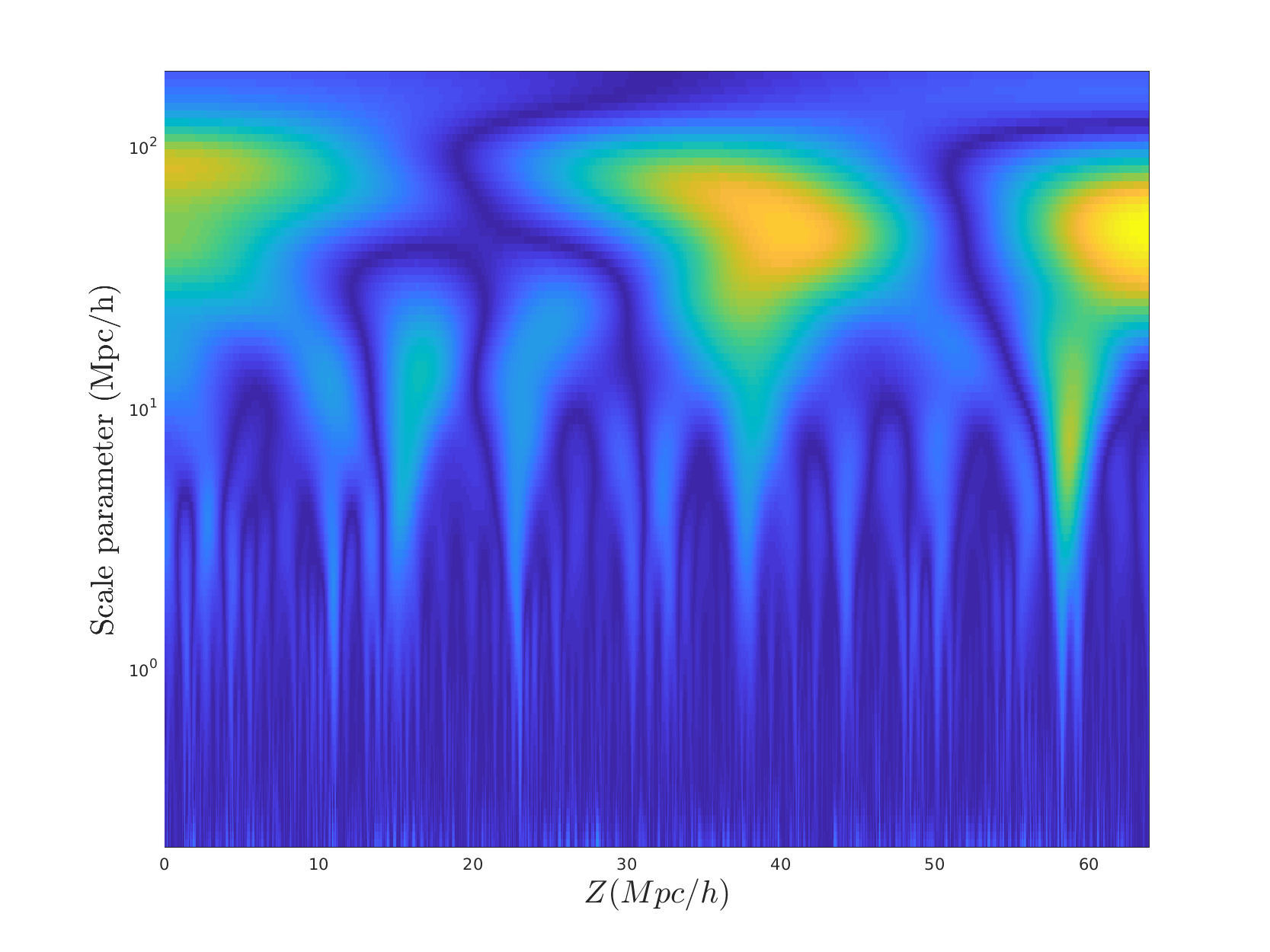}

\includegraphics[width=.5\textwidth]{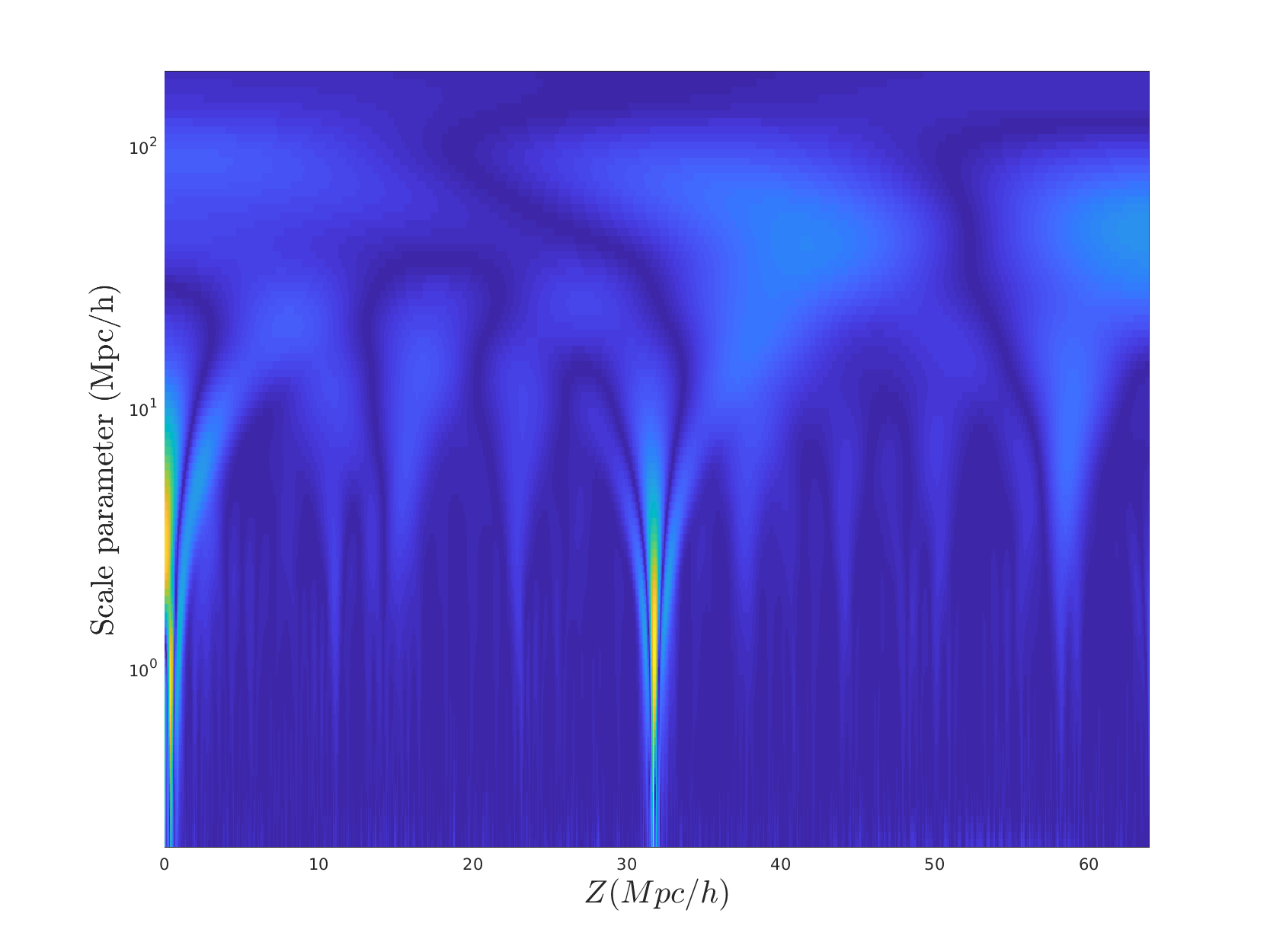}

\includegraphics[width=.5\textwidth]{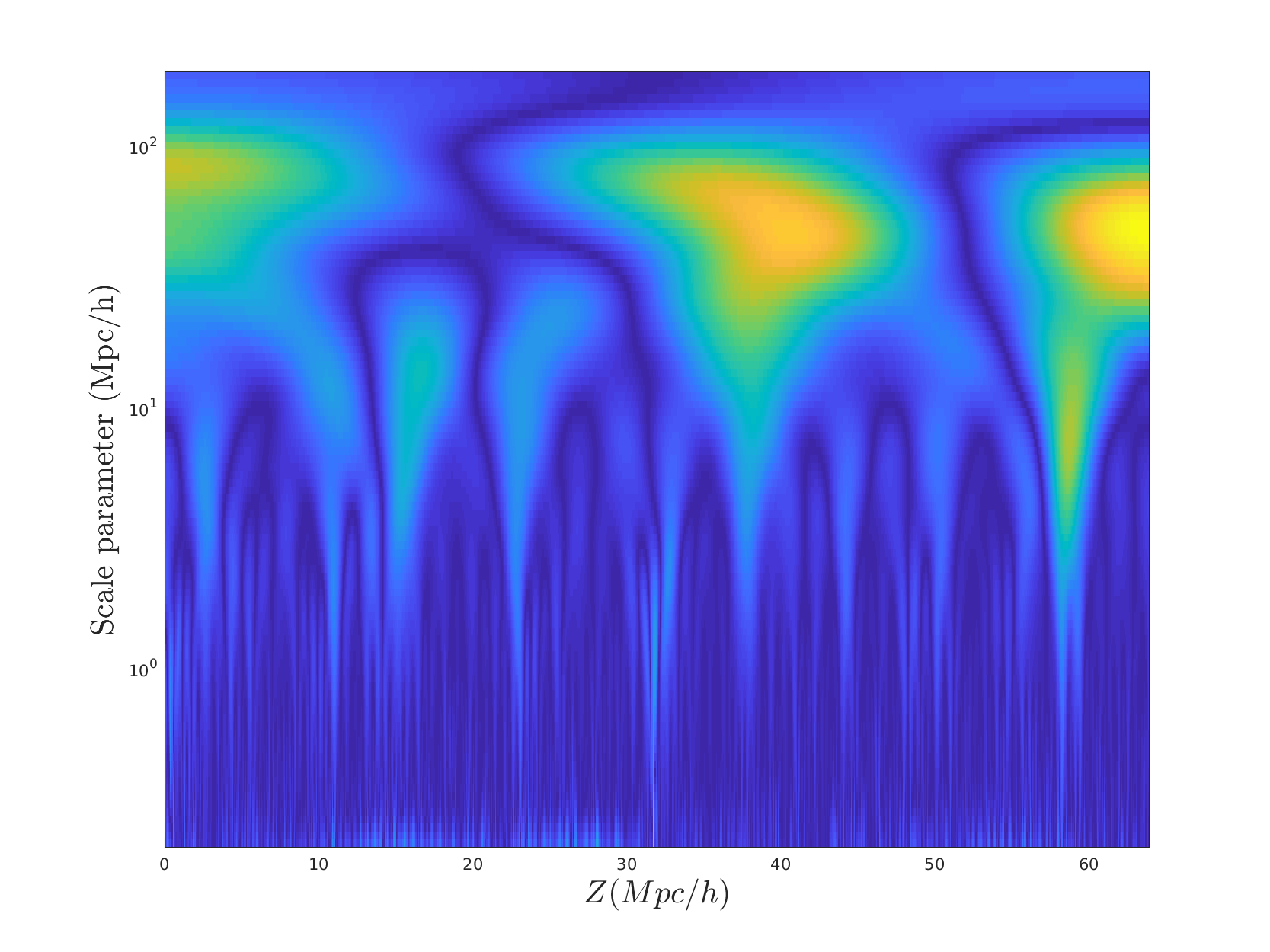}

\includegraphics[width=.5\textwidth]{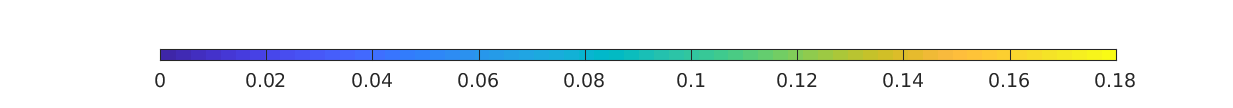}

\caption{Continuous wavelet transforms of the density contrast of the 1-d projected dark matter density.
The horizontal axis is the position parameter of the wavelet decomposition (with $Z = 32$ being the
position of the wake), and the vertical axis is the width of the wavelet. The top panel is for a simulation
without a wake, the middle panel includes a wake with $G\mu = 8 \times 10^{-7}$, and the bottom
panel has a wake with $G\mu = 10^{-7}$. Note that the $\Lambda$CDM fluctuations are the same in all
three simulations.}

\label{Fig7}

\end{figure}

We have applied the {\it continuous Morse wavelet transformation} \cite{Morse}
to the above 1-d projection plots, and below we show some of
the results. The basis used for the continuous wavelet transformation
is the Generalized Morse Wavelet \cite{Morse} which has two parameters: 
$\beta$, measuring compactness and $\gamma$, characterizing the symmetry of the 
Morse wavelet. We choose $\gamma=3$ (corresponding to the symmetric case). 
There are few oscillations if we choose $\beta$ close to $\gamma$, so 
$\beta=3.01$  is suitable for discontinuities detection. The wavelets 
are thus characterized  by the position $Z$
where they are centered, and by their scale parameter (width) $w$. 
In the following plots of figure \ref{Fig7}), the horizontal axis is $Z$, the vertical axis is the 
scale parameter. The color is a measure of the modulus of the
wavelet coefficients.

A wake is a very thin feature at a particular value of $Z$. Hence, the
wave signal will be concentrated at the lowest values of the scale
parameter. Figure 9 shows a comparison between the
wavelet transformation coefficients in simulations without a wake
(top) and including a wake with $G\mu = 10^{-7}$ (bottom)
and $G\mu = 8 \times 10^{-7}$ (middle), all at a redshift of $z = 10$.
The Gaussian noise gives rise to features in the continuous wavelet
map which are mostly concentrated at large scale parameters, although
there are some features which also appear at small scale parameters. As
seen in the bottom panel of Figure 7, the wake adds a narrow feature at the value of $Z$
where the wake is centered which continues to $w \simeq 0.4 h^{-1} {\rm Mpc}$. It can be
characterized as a narrow spike. The wake-induced spike and the
spike-like features in the no-wake simulation can be distinguished in
that the features coming from the Gaussian noise weaken as $w$ approaches
its minimum value, and are wider than the wake-induced spike. Note that the
color scaling is the same in the three panels. We see that the wake
signal stands out very strong at $z = 10$ for a wake with $G\mu = 10^{-7}$,
and that it totally dwarfs all other features for $G\mu = 8 \times 10^{-7}$.
The above maps are obtained for a high resolution sampling along the $Z$
direction (we move the center position of the wavelet in steps of $0.5$ of
the grid size).


Since the wavelet expansion is an expansion in a complete set of functions, it is
possible to reconstruct the original data from the wavelet transform. 
By setting to zero all wavelet coefficients 
corresponding to scales higher that a given cutoff, here taken as $0.4 h^{-1} {\rm Mpc}$, 
we can construct filtered 1-d projection graphs in which the long wavelength 
contributions of the Gaussian noise are eliminated and in which the string wake 
signal is more clearly visible. If we apply
the reconstruction algorithm to the filtered wavelet maps, we can construct a 
{\it filtered 1-d projection} graph in which the long wavelength contributions of the
Gaussian noise are eliminated and in which the string wake signal is more clearly
visible.

Figure 8 show the reconstructed filtered 1-d projection graphs at redshift
$z = 10$ in the case of pure Gaussian noise (top panel), and including a string wake
with $G\mu = 10^{-7}$ (bottom panel) and $G\mu = 8 \times 10^{-7}$ (middle panel). As in the
previous two graphs, the horizontal axis is the coordinate $Z$, and here the vertical axis is
the density contrast. The wake signals are greatly enhanced compared to what can
be seen in the unfiltered projection graphs. For the value of $G\mu = 10^{-7}$, the
wake signal is now almost an order of magnitude higher in amplitude than the peak
value in the case of pure $\Lambda$CDM fluctuations.

\begin{figure}

\includegraphics[width=.5\textwidth]{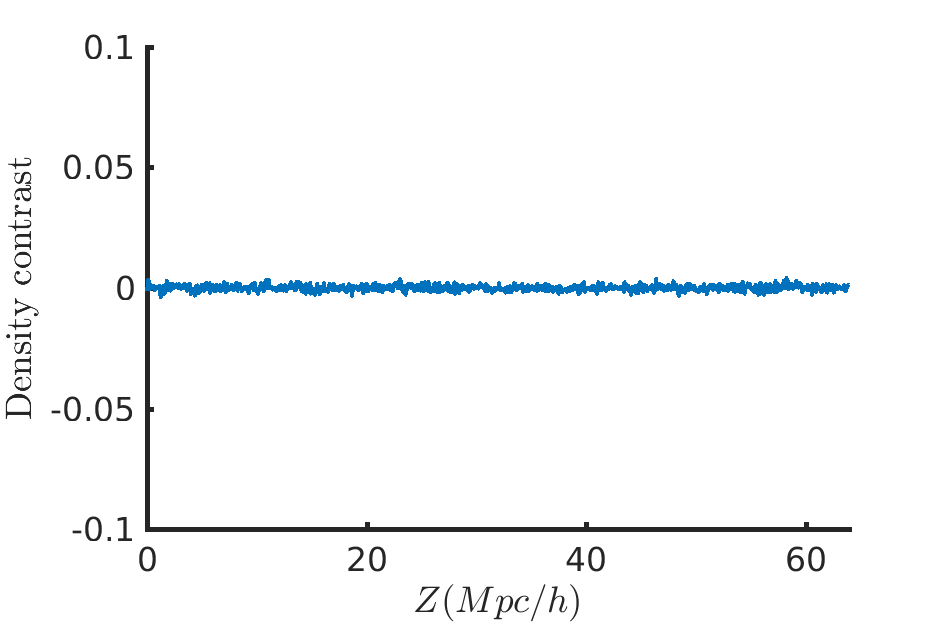}

\includegraphics[width=.5\textwidth]{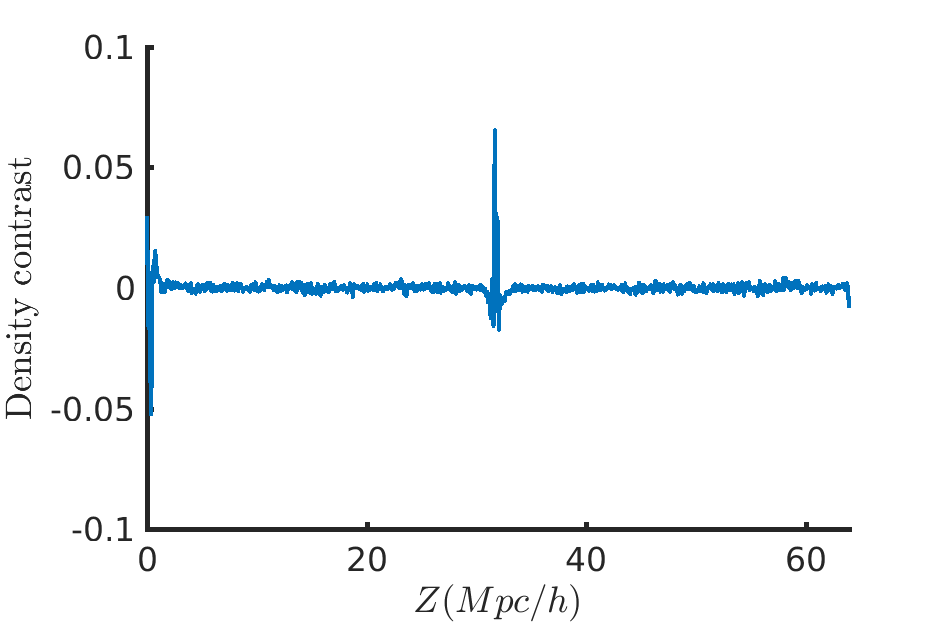}

\includegraphics[width=.5\textwidth]{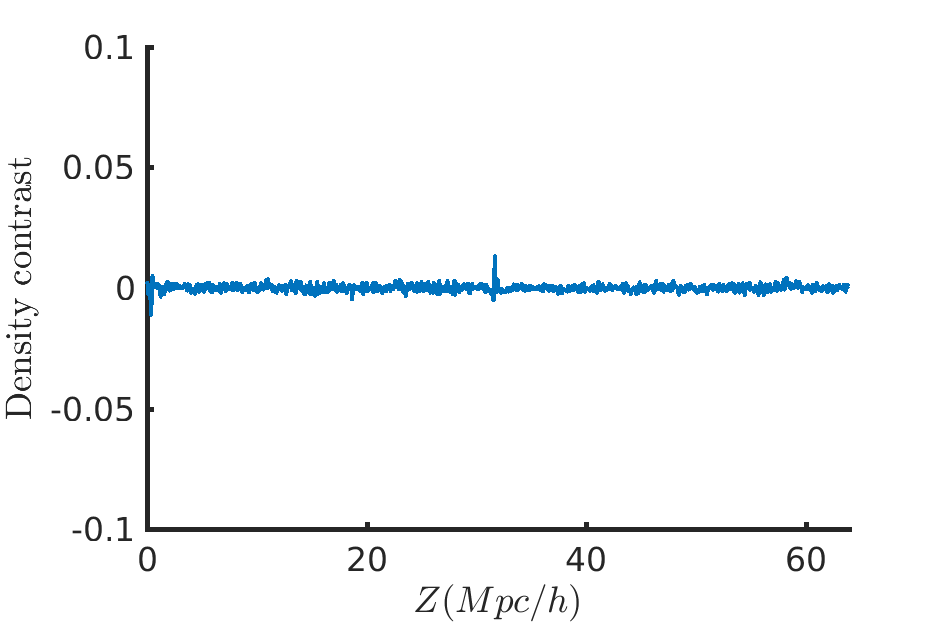}

\caption{Reconstruction of the 1-d density contrast from the filtered wavelet transforms. The vertical
axis is the density contrast, the horizontal axis is the distance of the projection plane from the wake plane.
The top panel is the result of a simulation without a wake, the middle panel has a wake with 
$G\mu = 8 \times 10^{-7}$, and the bottom panel has a wake with $G\mu = 10^{-7}$. The data
is for a redshift $z = 10$. Note that the wake signal is greatly enhanced compared to the original
1-d projection graphs of Fig. 8.}

\label{Fig9}

\end{figure}

To show a comparison, we can apply the continuous wavelet transformation to the one-dimensional projection filtered map, 
obtaining Figure 9. Comparing Figures 9 and 7 we see that the wake signal has been greatly 
enhanced by filtering.

\begin{figure}

\includegraphics[width=.5\textwidth]{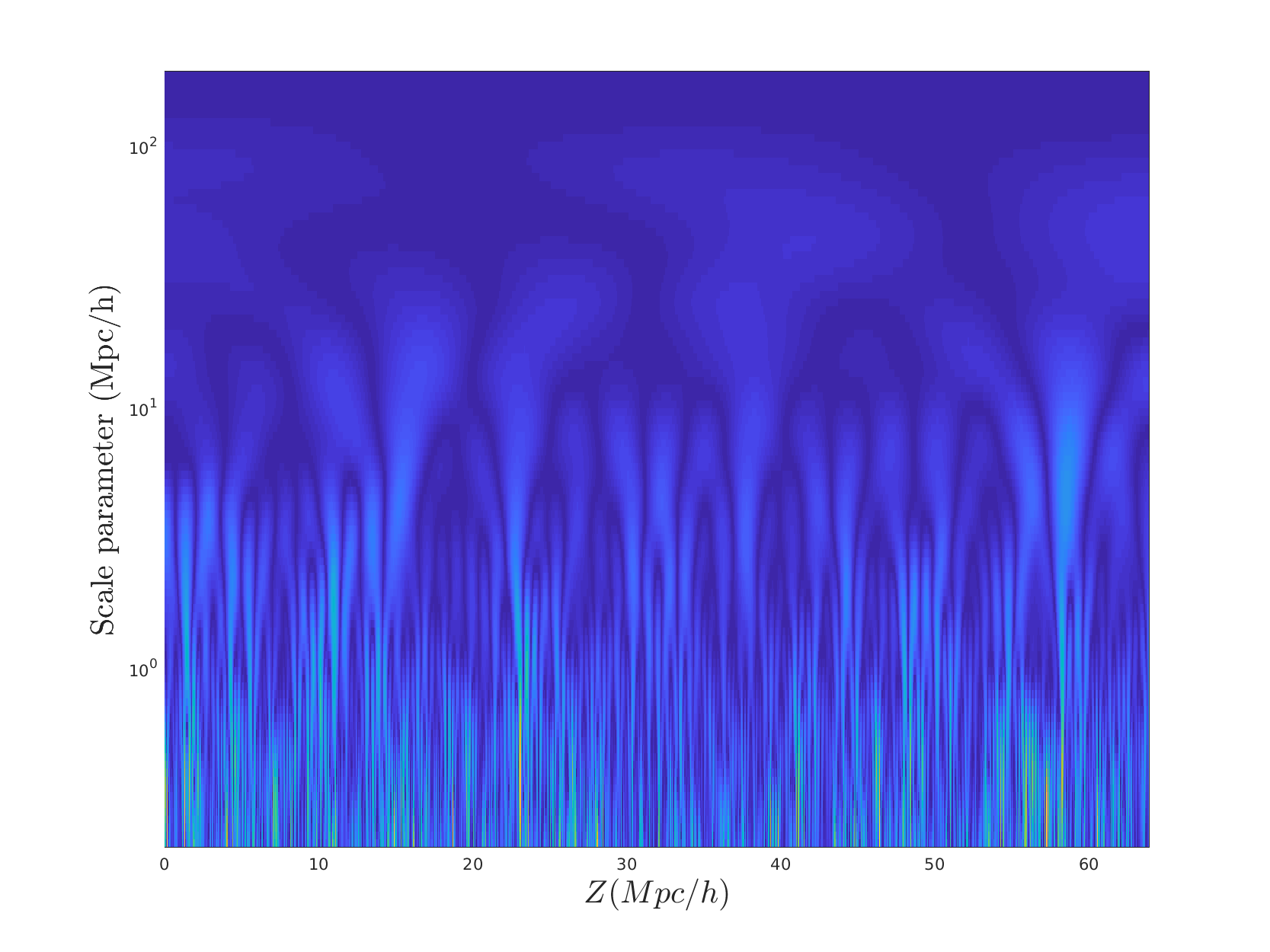}

\includegraphics[width=.5\textwidth]{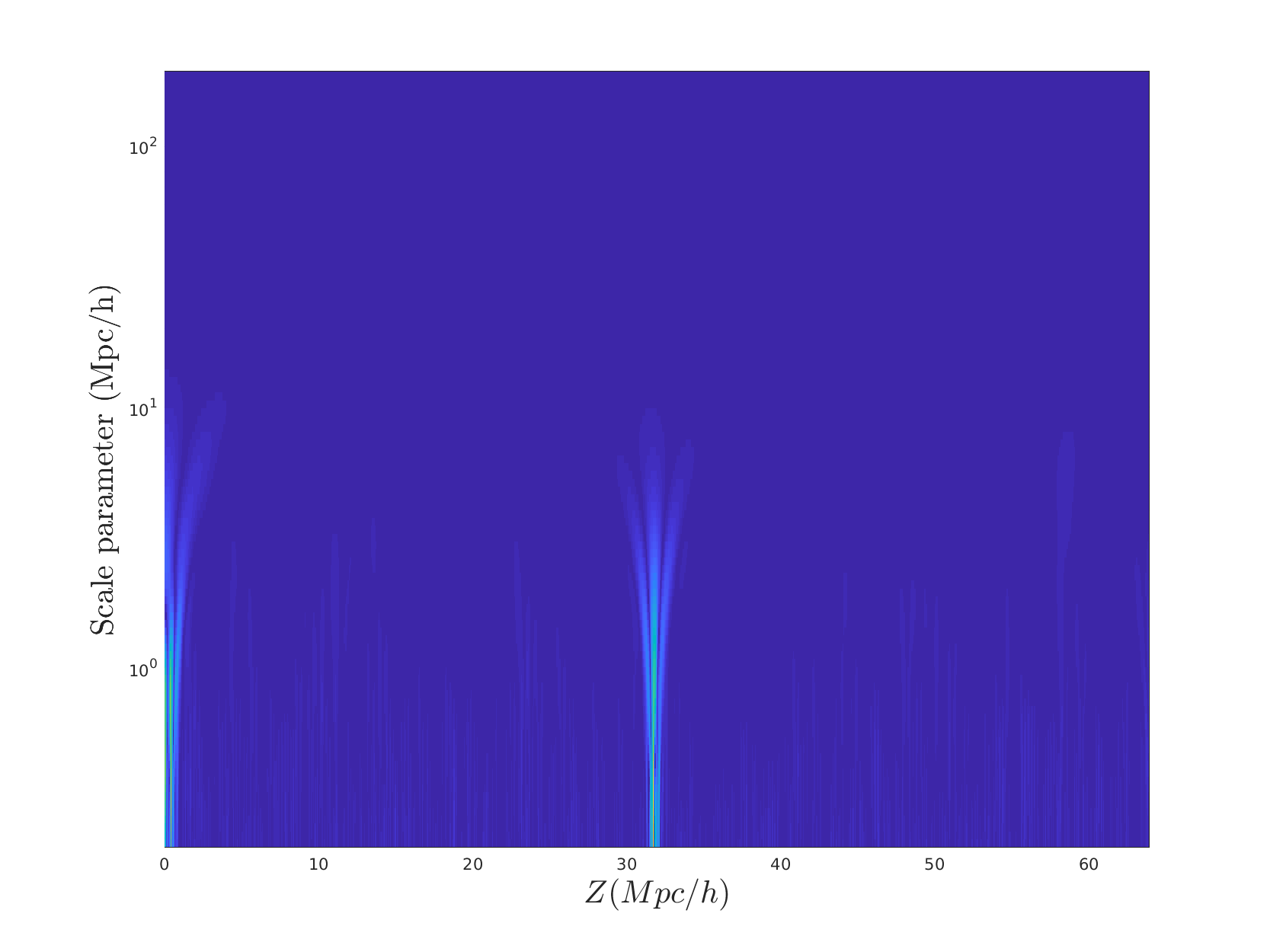}

\includegraphics[width=.5\textwidth]{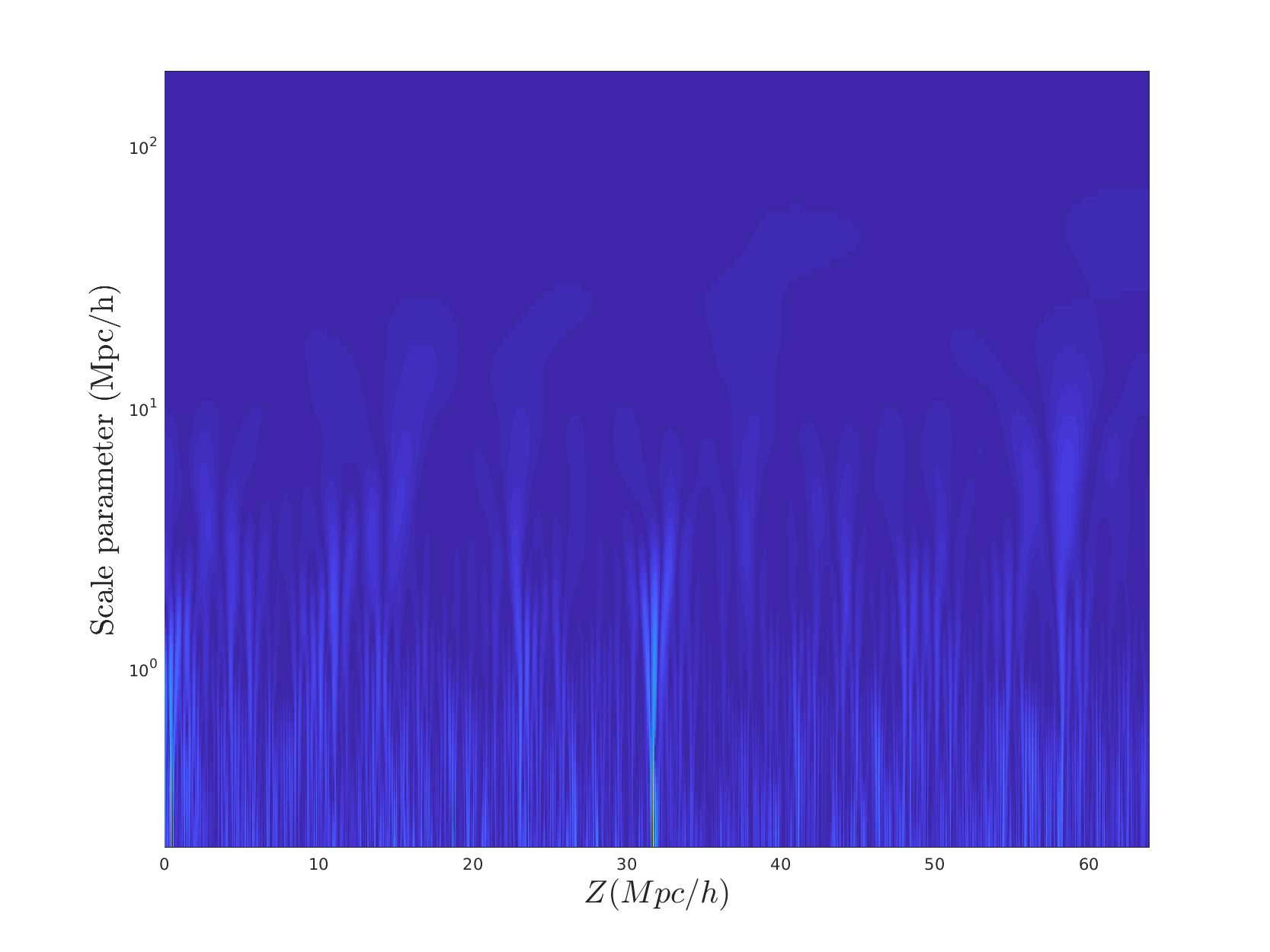}

\includegraphics[width=.5\textwidth]{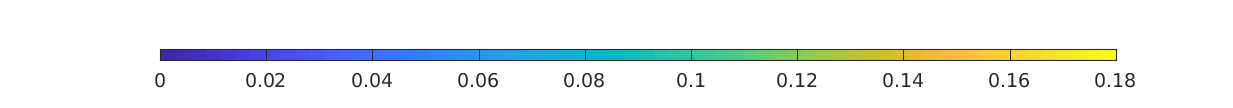}

\caption{Continuous wavelet transforms after filtering. The axes are like in the figure 7. The top panel
is without a wake, the middle panel has a wake with $G\mu = 8 \times 10^{-7}$, and the bottom panel
is for a wake with $G\mu = 10^{-7}$.}

\label{Fig10}

\end{figure}

A statistic which can be used to quantify the significance of the wake signal is the
{\it signal to noise ratio} ${\cal R}$ which we define to be
\be
{\cal R} \, = \, \frac{s\ -\bar{n}}{\sigma(n)} \, ,
\ee
where $s$ is the peak value of the filtered 1-d projection graph with a wake, and $\bar{n}$ is the mean of the peak value of the filtered 1-d projection graph without a wake for the 10 samples and $\sigma(n)$ is the standard deviation of the peak value of the filtered 1-d projection graph without a wake for the same 10 samples. 

We find that the average of the signal to noise ratio is
\be
{\cal \bar{R}} \, = \, 17.07 \pm 0.83 \, ,
\ee
at redshift $z=10$ in the case of a wake with $G\mu = 10^{-7}$ initially laid down at redshift $z=15$ (the error bars are the standard deviation
based on three simulations). 
Hence, we find that
a cosmic string wake is identifiable 
with a $ 17.07 \sigma$ significance. At redshift $z = 7$ the difference in the signal
to noise is no longer statistically significant.

The results of the signal to noise analysis are given in the following figure (10). The horizontal axis gives the redshift (early
times on the left), and the vertical axis is the signal to noise ratio. The bottom
curve gives the results for a pure $\Lambda$CDM simulation, the next pair
of curves (counting from bottom up) give the results of a simulation where
a wake with $G\mu = 10^{-7}$ is added, and the top two curves correspond
to adding a wake with $G\mu = 8 \times 10^{-7}$. The two members of the
pair correspond to different redshifts of inserting the wake. The difference
in the predictions by changing the wake insertion time is not statistically
significant.This figure shows the signal to noise ratios of the reconstructed
1-d projection graphs after wavelet transformation and filtering, for the high sampling scale.
For $G\mu = 10^{-7}$ the wake effect can be clearly seen up to
a redshift of $z = 10$, and for $G\mu = 8 \times 10^{-7}$ up to a redshift
of $z = 5$.

\subsection{Spherical Statistic}

In the previous analysis, the statistical analysis was based on an algorithm which
presupposed knowing the planar orientation of the wake. For applications to data,
we need an analysis tool which does not use this information. In this section we
develop such a statistic, which is an adaptation of a 3D ridgelet analysis (\cite{Starck}).

The idea is to take the filtered one dimensional projection, similar to Figure 8, for 
different directions on the sky and compute a relevant quantity. For choosing
the angles, a Healpix \cite{Gorski:2004by} scheme was used \footnote{we choose $Nside=512$, }  
with the help of S2LET  \cite{Leistedt} , a free package available at www.s2let.org.
The statistic is constructed in the following way: for any direction of the sphere, we consider an 
associated projection axis passing through the origin of the box. 
We then consider slices of the simulation box perpendicular to that axis at each point x of it with 
thickness given by the grid size of the simulation, and we compute the mass density $\delta (x)$ 
of dark matter particles in that slice as  function of x. The range of x is half the simulation box 
so we avoid slices with small area (compared with the face of the cubic simulation box). 
A one-dimensional filter wavelet analysis similar to the one described in the previous 
section is then performed in the mass density $\delta (x)$, giving a filtered version for it, 
called $f\delta$.
 
We then compute the maximum value S of $f\delta (x)$ for each direction on the simulation.
S is the map on the surface of the sphere which we now consider and $\hat{S}$ is its maximum value. 
For a simulation including a cosmic string wake with $G\mu = 1\times 10^{-7}$ the resulting map at redshift
z = 10 is shown in the top left panel of Figure 11, whereas the analysis without the wake is shown 
in Figure 12. The value of S is indicated in terms of color (see the sidebar for the values). 
 
The wake signal appears at the center of the box and its associated S value is about $40\%$ 
higher than the maximum of the S value for the map without wake. This can be better visualized in figures 13 and 14, where the center in zoomed in $40$ times. For each map, a peak 
$\hat{S}$  over standard deviation $\sigma$ of $S$ was computed, and it was found that 
$\hat{S}/ \sigma = 8.68 $ for a simulation without wake and $\hat{S}/ \sigma = 14.32 $ for 
the simulation with a $1\times 10^{-7}$ wake. 

A wake perpendicular to a particular direction will yield a high signal
since the mass in the slice which overlaps with the wake will get a
large contribution localized at a particular value of x. 

The Healpix scheme does not contain the angles $(\theta,\phi)=(0,0)$ where we know the 
wake is located, so we never probe the orientation exactly were the wake is. The wake signal 
has to be reached by increasing the resolution of the analyzed angles. This supports 
the idea that the information regarding the orientation of the angles should not be included 
in the analysis.

\begin{widetext}

\begin{figure}[H]
\includegraphics[height=8cm]{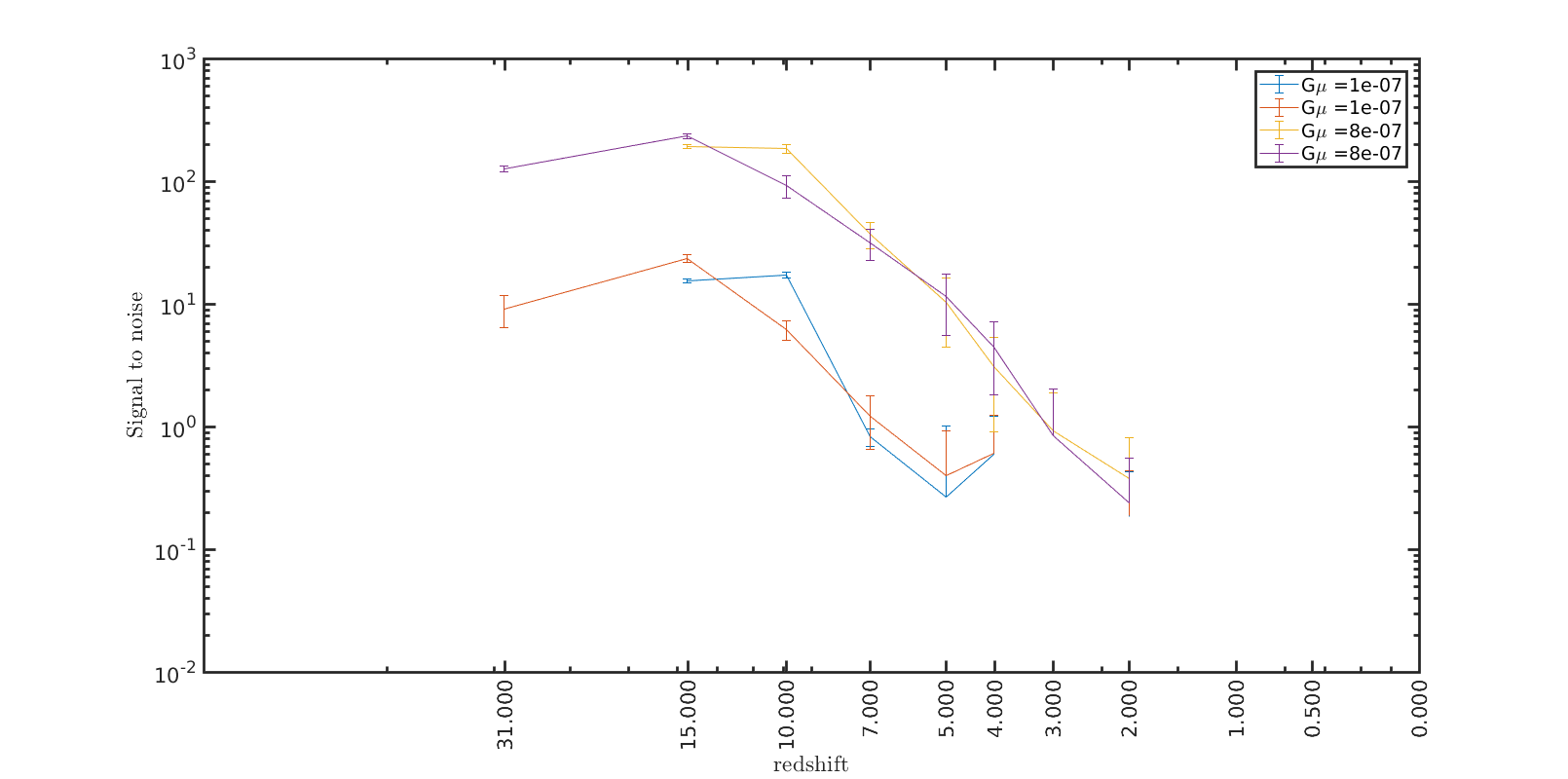}
\caption{Signal to noise analysis for the 1-d projections after wavelet
transformation, filtering and reconstruction, and for the finer of samplings in wavelet width. The horizontal
axis shows the redshift, the vertical axis is the signal to noise ratio. The two pair of curves in the bottom contain
an wake with $G\mu = 10^{-7}$, and the top two curves are for an
wake with $G\mu = 8 \times 10^{-7}$. The two members of a pair of curves 
correspond to different redshifts of wake insertion (as is obvious from the
starting points of the curves). The points that are not shown correspond to values for the signal to noise equal to zero.} 
\label{}
\end{figure}

\end{widetext}

\begin{widetext}

\begin{figure}[h]
\includegraphics[width=0.8\textwidth]{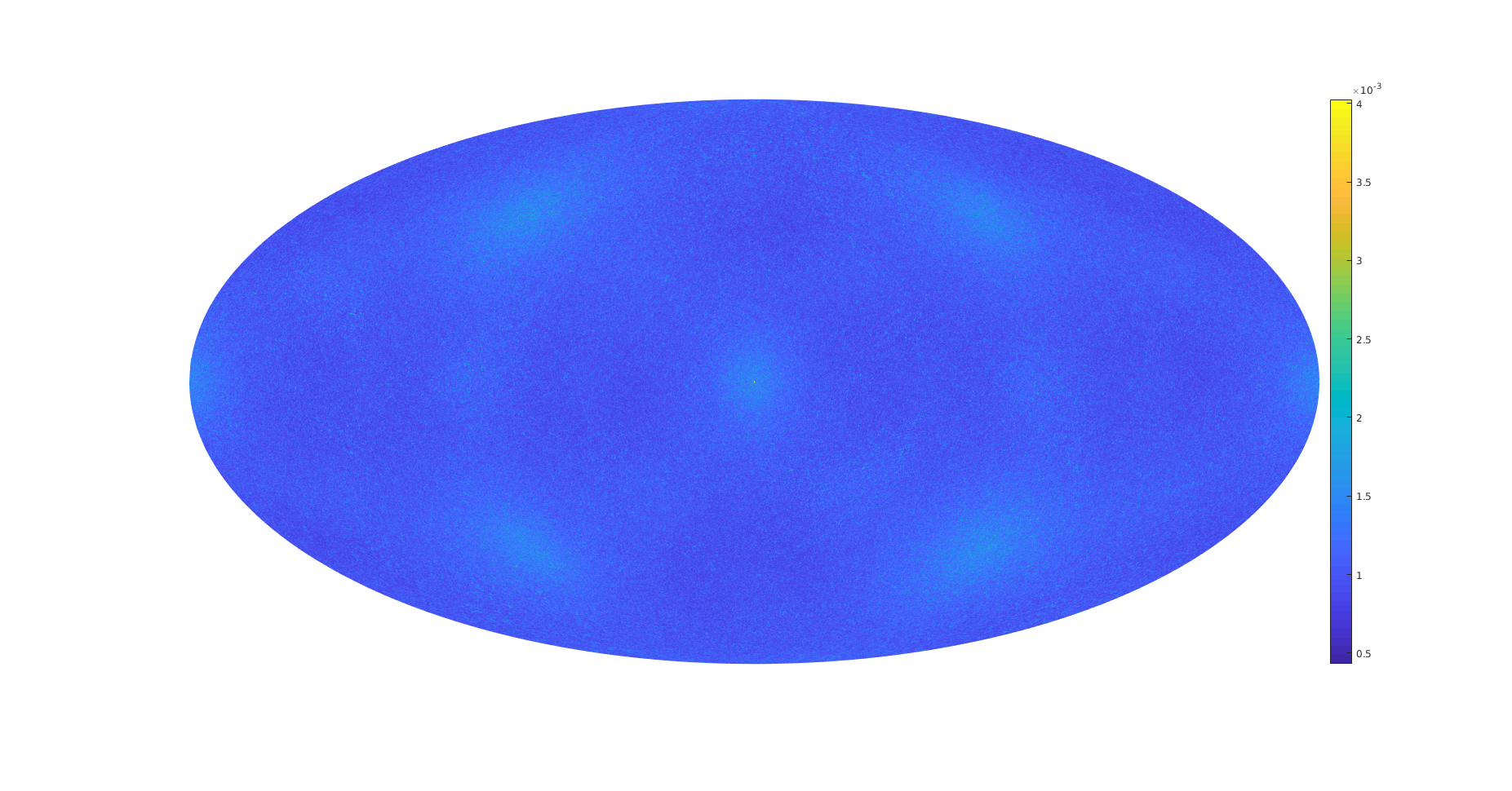}
\caption{Reconstructed map for a simulation with wake tension $G \mu = 1 \times 10^{-7}$
at redshift $z = 10$.}

\includegraphics[width=0.8\textwidth]{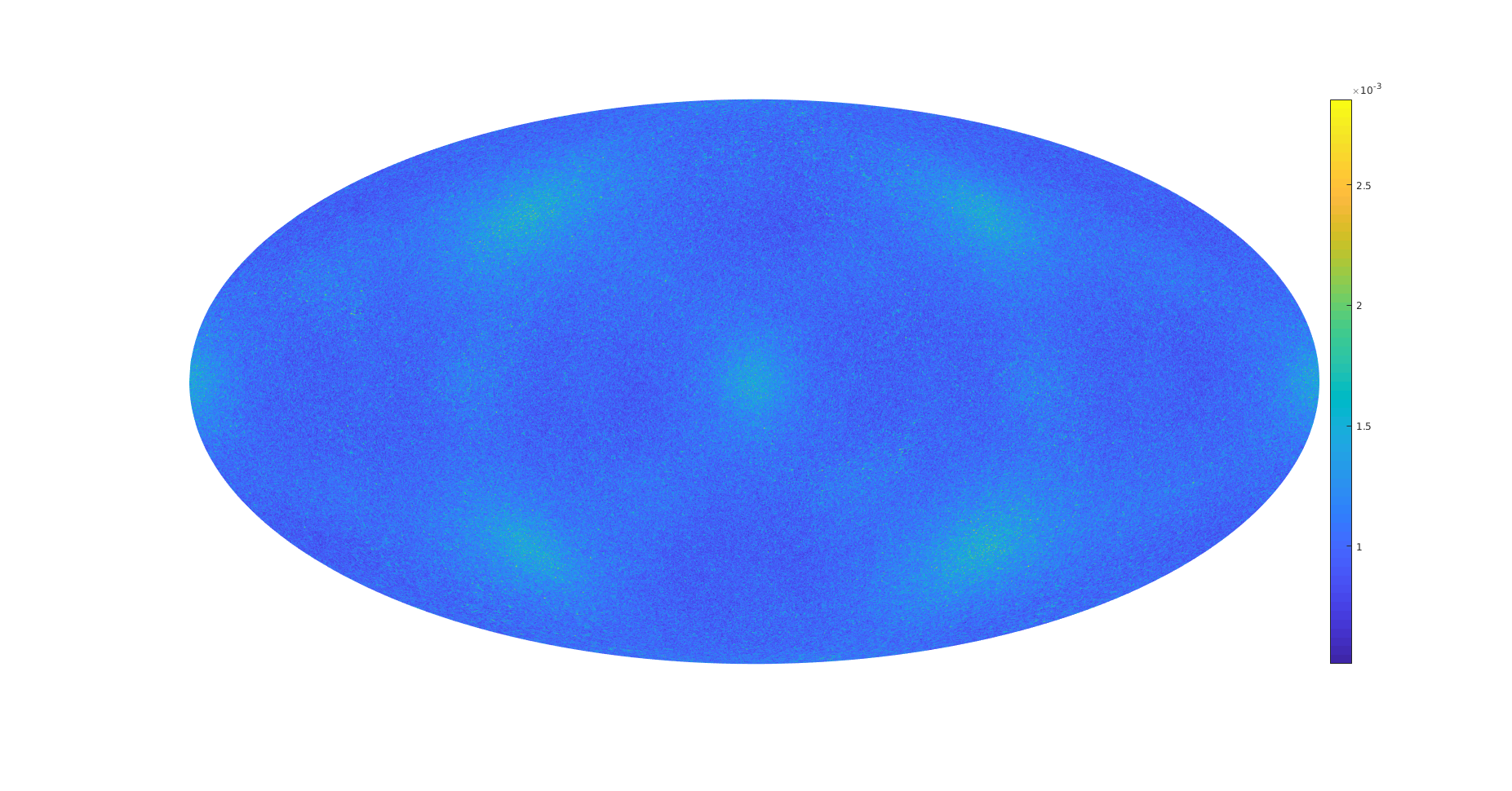}
\caption{The same map for a simulation without a wake, at redshift $z = 10$.}
\end{figure} 
\end{widetext}

\begin{figure}
\includegraphics[width=0.4\textwidth]{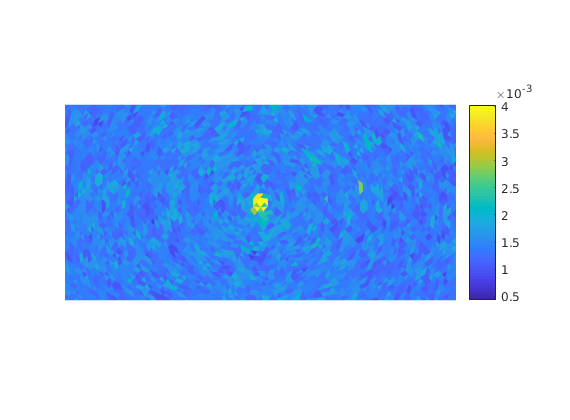}
\caption{$40\ \times$ zoom of the reconstructed map for a simulation with wake tension $G \mu = 1 \times 10^{-7}$
at redshift $z = 10$.}

\includegraphics[width=0.4\textwidth]{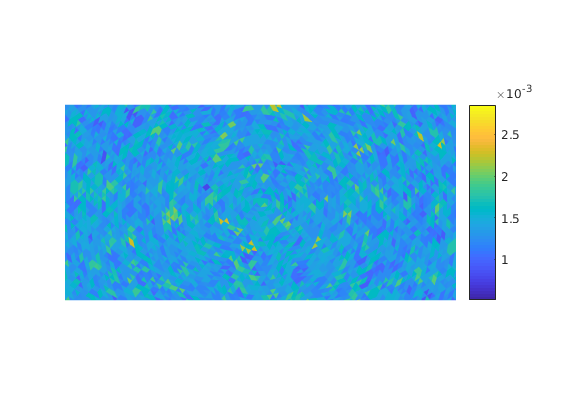}
\caption{$40\ \times$ zoom of the the same map for a simulation without a wake, at redshift $z = 10$.}
\end{figure} 

At the present level, our
analysis shows that cosmic string wakes with a tension
of $1 \times G\mu = 10^{-7}$ can be extracted at redshift $z = 10$,
as was found in our previous study where knowledge of the
orientation of the wake was assumed. In work in progress we
are investigating whether string wakes are in fact visible at
lower redshifts using this more sophisticated statistic. One
could first imagine that an analysis which uses the knowledge
of the wake orientation will yield better results than one which
does not, but this may not be the case here since the analysis
of this section uses more properties which differentiate $\Lambda$CDM
fluctuations and wake signals than the previous analysis did.

\section{Conclusions and Discussion}

We have presented the results from N-body simulations of the effects of a planar
cosmic string wake on the distribution of dark matter.
We have demonstrated that cosmic string signals can be extracted from
the background of $\Lambda$CDM fluctuations by considering the three dimensional
distribution of dark matter. Given the current resolution of the simulations, a threshold
of
\be
G\mu \, = \, 10^{-7}
\ee
can be reached if the dark matter distribution is considered at redshift $z = 10$. This
value of the string tension is competitive with the current limit which stems from
the angular power spectrum of CMB anisotropies. With improved resolution, improved
limits may be within reach. This means that the string signal might be identifiable 
for $G\mu = 10^{-7}$ even at redshifts lower than $z = 10$, and for smaller values
of $G\mu$ at redshift $z = 10$.

Key to this work is that we are looking for the specific non-Gaussian signals which
cosmic string wakes induce in position space. Position space algorithms are much
more powerful at identifying cosmic string signals than by focusing simply on the
power spectrum. It is possible that with improved statistical tools, better limits on
the string tension can be reached. Work on this question is in progress.

Searching for signals of individual cosmic string wakes in position space has a
further advantage compared to studying only the power spectrum: the position
space algorithms are to first approximation insensitive to the number $N$ of
strings passing through each Hubble volume. This number is known only to
within an order of magnitude, although we know from analytical arguments
(see e.g. \cite{RHBCSrev}) that $N$ should be of the order one. In particular,
this means that the constraint (\ref{CSbound}) on $G\mu$ from the angular power spectrum
of CMB anisotropies is sensitive to the value of $N$ which is assumed, whereas
our analysis is not.

In this work we have shown that by analyzing the distribution of dark matter, an
interesting threshold value of the cosmic string tension can be reached. In future
work, we plan to explore how changing the size of the simulation box, the spatial
resolution of the simulations, and the sampling width can lead to improved
bounds.

So far we have considered simulations with a single cosmic string wake. An
extension of our work will involve studying the effects of a full scaling distribution
of strings. This work will be conceptually straightforward but computationally
intensive. Another extension for the current work would be to consider curvelet-like signal extraction, in which segments of the wake would be detected, possibly giving more information about the wake presence when it is bended due the the $\Lambda CDM$ fluctuations at low redshifts and it is not a plane anymore.

In order to compare our simulation results to observational surveys, we need to
extend our work in several ways. To compare our work with optical and infrared
galaxy survey results, we need to identify halos from our distribution of dark matter,
and run the statistical tools on the resulting distribution of halos. The N-body
code we are using already contains a halo-finding routine. Hence, this extension
of our work will also be straightforward. 

As we have seen, the string signals are much easier to identify at higher redshifts.
Hence, 21cm surveys might lead to tighter constraints on the cosmic string
tension. At redshifts lower than the redshift of reionization, most of the neutral
hydrogen which gives 21cm signals is in the galaxies. Hence, the distribution
of 21cm radiation could be modeled by considering the distribution of galaxy
halos obtained from our simulations and by inserting into each halo the
distribution of neutral hydrogen obtained recently in the study of \cite{Hamsa}.
We plan to tackle this question in the near future.

The effect of cosmic strings on the 21cm signal from the {\it Epoch of Reionization}
is a more difficult question. Here, the ionizing radiation from cosmic string
loops (e.g. via cosmic string loop cusp decay \cite{cusp}) 
will most likely play a dominant role. Finally, at redshifts greater than that of
reionization, string wakes lead to a beautiful signal in 21cm maps: thin wedges
in redshift direction extended in angular directions to the comoving horizon
at $t_{eq}$ where there is pronounced absorption of 21cm radiation due
to the neutral hydrogen in the wake \cite{Holder2}.

\section*{Acknowledgement}
\noindent

The collaborative research between McGill University and the ETH
Zurich has been supported by the Swiss National Science Foundation
under grant \texttt{IZK0Z2\_174528}. In addition, the research at McGill has been supported in part by funds from the Canadian NSERC and from the Canada Research Chair program. DC wishes to acknowledge  CAPES (Science Without Borders) for
a student fellowship. JHD is supported by the European Commission under a Marie-Sklodowska-Curie European Fellowship (EU project 656869). This research was enabled in part by support provided by Calcul Quebec (http://www.calculquebec.ca/en/) and Compute Canada (www.computecanada.ca).

\appendix

\section{Peak Heights}

In this appendix we add a figure showing the distribution
of peak heights used in the statistic in Section IV.A.
The horizontal axis gives the redshift, the vertical axis
is the peak height. Shown are the results for ten
simulations without a wake (marked ``o") and for three
simulations each with a wake with tension $G\mu = 10^{-7}$
inserted at redshift $z = 31$ (marked ``*''), with the
same tension and insertion redshift $z = 15$ (marked ``+''),
with tension $G\mu = 8 \times 10^{-7}$ and insertion
redshifts $z = 31$ (marked with inverted
triangles) and $z = 15$ (marked ``x'').

\begin{widetext}

\begin{figure}[H]
\includegraphics[width=0.9\textwidth]{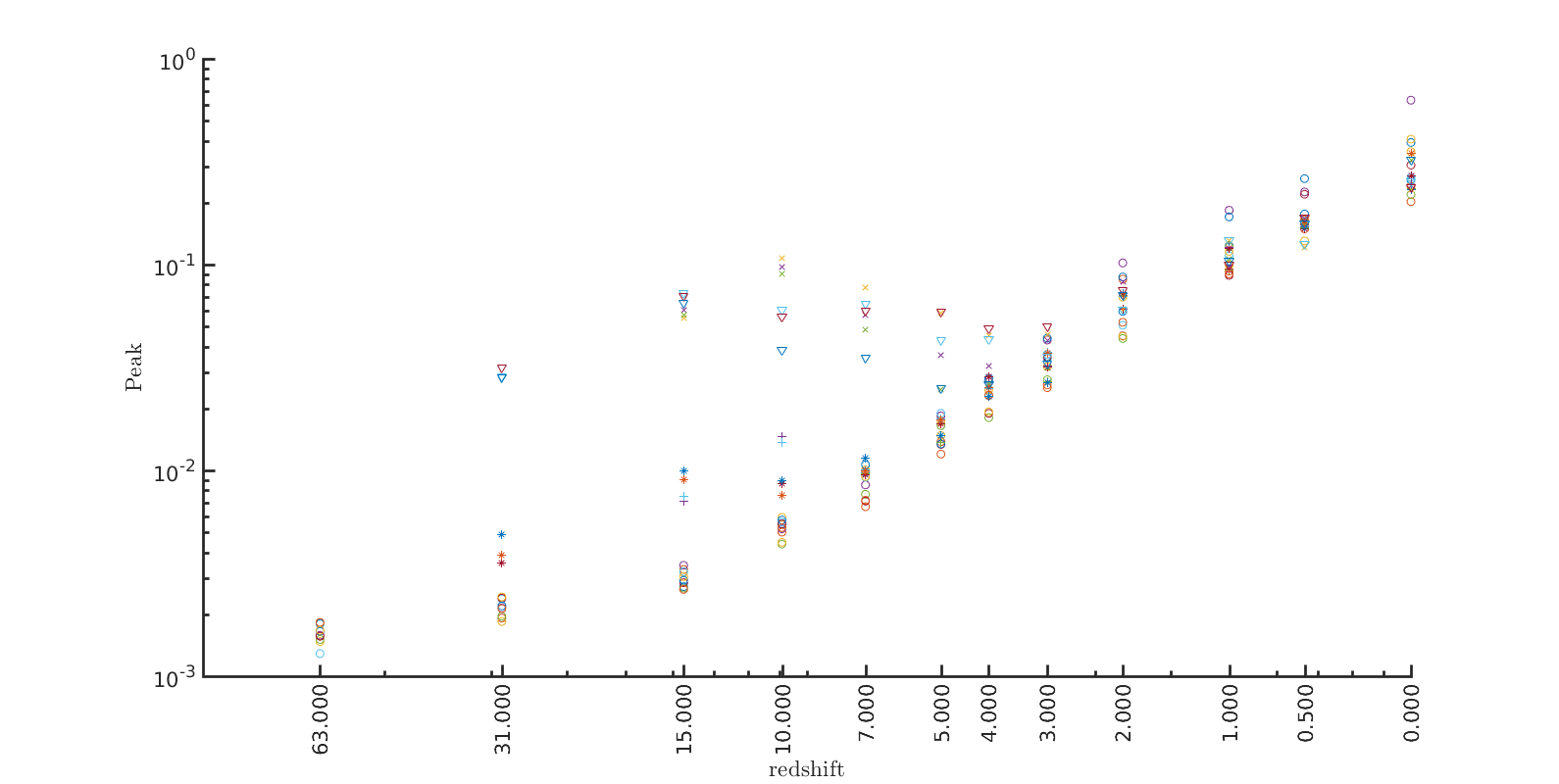}
\caption{The peak values for simulations without a wake are marked
by circles, for those with a wake with tension
$Gμ = 10^{−7}$ inserted at redshift $ z = 31$
and $z = 15$ by ``*" and ``+", respectively, and for those with
tension $Gμ = 8 \times 10^{−7}$ 
inserted at redshifts $z = 31$ and $z = 15$ by inverted 
triangles and ``x", respectively.}
\end{figure} 
\end{widetext}

From this figure it is clear that wakes with $G\mu = 10^{-7}$
are identifiable until redshift of $z = 10$ while those
with $G\mu = 8 \times 10^{-7}$ can be identified down to
below $z = 7$.


\end{document}